\documentclass[journal]{IEEEtran}
\usepackage{graphicx}
\usepackage{subfigure}
\usepackage{float}
\usepackage{subfloat}
\usepackage{amsmath}
\usepackage{amssymb}
\usepackage{amsthm}
\usepackage{epstopdf}
\usepackage{cases}
\usepackage{color}
\usepackage{makecell}
\usepackage{cite}
\usepackage{algorithmic}
\usepackage[ruled]{algorithm2e}

\DeclareMathOperator{\arccot}{arccot}

\newtheorem{theorem}{Theorem}

\usepackage[normalem]{ulem}

\ifCLASSINFOpdf
\else
\fi
\usepackage{caption}
\usepackage{mathtools}

\begin{document}

%
\title{Sensing Performance of Cooperative Joint Sensing-Communication UAV Network}
%
%
%

%

\author{Xu Chen,~\IEEEmembership{Student Member,~IEEE,}
	Zhiyong Feng,~\IEEEmembership{Senior Member,~IEEE,}
	Zhiqing Wei,~\IEEEmembership{Member,~IEEE,}
	Feifei Gao,~\IEEEmembership{Fellow,~IEEE,}
	and Xin Yuan,~\IEEEmembership{Member,~IEEE}
	\thanks{Part of related work was accepted by IEEE International Conference on Signal, Information and Data Processing (ICSIDP) 2019 \cite{Chen2019Per}. Xu Chen, Z. Feng, Z. Wei and Xin Yuan are with Beijing University of Posts and Telecommunications, Key Laboratory of Universal Wireless Communications, Ministry of Education, Beijing 100876, P. R. China (Email:\{chenxu96330, fengzy, weizhiqing, yuanxin\}@bupt.edu.cn).}
	\thanks{F. Gao is with Institute for Artificial Intelligence, Tsinghua University (THUAI), State Key Lab of Intelligent Technologies and Systems, Tsinghua University, Beijing National Research Center for Information Science and Technology (BNRist), Department of Automation, Tsinghua University Beijing, P. R. China (email: feifeigao@ieee.org). }
	\thanks{Corresponding author: Zhiyong Feng, Zhiqing Wei}
}

%
%

\markboth{}%
{Shell \MakeLowercase{\textit{et al.}}: Bare Demo of IEEEtran.cls for IEEE Journals}
%



\maketitle

\pagestyle{empty}  
\thispagestyle{empty} 

\begin{abstract}
We propose a novel cooperative joint sensing-communication (JSC) unmanned aerial vehicle (UAV) network that can achieve downward-looking detection and transmit detection data simultaneously using the same time and frequency resources by exploiting the beam sharing scheme. The UAV network consists of a UAV that works as a fusion center (FCUAV) and multiple subordinate UAVs (SU). All UAVs fly at the fixed height. FCUAV integrates the sensing data of network and carries out downward-looking detection. SUs carry out downward-looking detection and transmit the sensing data to FCUAV. To achieve the beam sharing scheme, each UAV is equipped with a novel JSC antenna array that is composed of both the sensing subarray (SenA) and the communication subarray (ComA) in order to generate the sensing beam (SenB) and the communication beam (ComB) for detection and communication, respectively. SenB and ComB of each UAV share a total amount of radio power. Because of the spatial orthogonality of communication and sensing, SenB and ComB can be easily formed orthogonally. The upper bound of average cooperative sensing area (UB-ACSA) is defined as the metric to measure the sensing performance, which is related to the mutual sensing interference and the communication capacity. Numerical simulations prove the validity of the theoretical expressions for UB-ACSA of the network. The optimal number of UAVs and the optimal SenB power are identified under the total power constraint.

\end{abstract}


%
\IEEEpeerreviewmaketitle

\section{Introduction}
%
%
%
%

Cooperative sensing unmanned aerial vehicle (UAV) network has been very promising in disaster rescue, surveillance, resource exploration, etc \cite{teacy2010maintaining}. Due to unmanned feature and power restriction, UAVs need to achieve functions such as communication, environment sensing and flying status perception under load, volume and spectrum constraints. The utilization of the joint sensing-communication (JSC) technique on the UAV platform is a reasonable choice for the cooperative sensing UAV network, because it has advantages in load saving and spectrum reuse by sharing the same antennas, transceivers and spectrum to achieve sensing and communication \cite{han2013joint}. With the integration of the sensing data from multiple UAVs using the JSC techniques, the sensing zone of the network will be extended far beyond the radio propagation limit of single sensing platform \cite{Sturm2011Waveform}. Thus, with limited spectrum for radar sensing and communication, the cooperative JSC UAV network can finish detecting a large area in much shorter time and gather more environment information than an individual UAV does, which makes the network be more agile and decide on a larger picture. 

The achievement of the JSC technique gains increasingly sufficient fundamentals. The communication spectrum has gradually occupied the  frequency band that was dedicated to radar at first~\cite{Sturm2011Waveform}. Digital signal processing techniques have been widely utilized both in sensing and communication transceivers \cite{Sturm2011Waveform, gupta2015survey}. Thus, the spectrum and hardwares that are separately designed for sensing and communication have great potential for convergence \cite{han2013joint}.

The strong need and potential for the joint design of communication and sensing has motivated a number of important studies in the JSC technique. The approaches to realize the JSC system can be divided into three categories: time sharing, waveform co-design and beam sharing. Time sharing scheme has the intrinsic disadvantage that it does not allow simultaneous operation of sensing and communication, which will lead to the loss of detection target \cite{moghaddasi2016multifunctional}. As for waveform co-design scheme, the communication direction is rigidly constrained by sensing direction. Therefore, the freedom of communication is severely reduced \cite{Wu8168372,shi2018power}. On the contrary, the beam sharing scheme allows simultaneous operation of sensing and communication using different beams, which can carry out communication and sensing as freely as possible \cite{liu2018mu, mccormick2017simultaneous, richards2005fundamentals,Chen2019Per}. The beam sharing scheme is based on the beamforming technology that has solid fundamentals. Monzingo and Miller proposed maximum signal-to-noise ratio (SNR) beamforming to generate optimal beam with the existence of strong noise\cite{monzingo2004introduction}. Frost proposed linearly-constrained minimum variance beamforming that can generate the beamforming vector with linear complexity \cite{frost1972algorithm}. Shi and Feng proposed a two-step iterative beamforming algorithm to achieve beamforming of high performance \cite{shi2005new}. Another fundamental aspect in JSC implementation is the JSC waveform. Traditional sensing waveform design concentrates on the waveforms with desirable autocorrelation properties. Linear frequency modulated (LFM) pulse signal, i.e., ``chirp'' signal, is a typical sensing waveform \cite{skolnik1970radar}. Exploiting the  high SNR characteristics of radar signal, a bio-inspired radio frequency (RF) steganography scheme that can conceal digital communication in linear chirp radar signals was proposed in  \cite{zhang2016bio}. Frequency modulated continuous wave (FMCW) is the typical  continuous-wave sensing waveform \cite{stove1992linear}. In addition, the direct sequence spread spectrum (DSSS) waveform is also widely studied due to its features in information security and spectrum spread gains \cite{nathanson1991radar}. C. Sturm proposed an orthogonal-frequency-division-multiplexing (OFDM) symbol-based sensing processing technique to take advantage of temporal and frequency domain signal \cite{sturm2010performance}. 

However, all the above works only study a single JSC unit or a single pair of JSC. Thus, in \cite{Chen2019Per}, we proposed a cooperative sensing performance metric of the JSC UAV network based on beam sharing scheme, but we were unable to put forward the proper antenna array model and beamforming algorithm. Moreover, we simplified the model by neglecting the mutual radar sensing interference and the guard distance between UAVs, which renders the proposed sensing performance metric less practical. Thus, in this paper, to enhance the concept of beam sharing JSC cooperative sensing UAV network we proposed in \cite{Chen2019Per}, we focus on the practical implementation of beam sharing JSC UAV network and further model the mutual sensing interference to analyze the sensing performance of cooperative JSC UAV network, which will offer guidance to the optimal deployment of the JSC UAV network, such as power allocation and the size of UAV network. The JSC UAV network consists of a UAV that works as the fusion center (FCUAV) and multiple subordinate UAVs (SU). All UAVs fly at the fixed height. UAVs carry out downward-looking sensing and integration of sensing data by generating a sensing beam (SenB) and a communication beam (ComB) with a novel JSC antenna array, respectively. The proper beamforming algorithm is also proposed. Because of the spatial orthogonality of communication and sensing, SenB and ComB can be easily formed orthogonally. SenB and ComB of each UAV share a total amount of available radio power. The average mutual sensing interference is modeled based on the radio propagation theory. The upper bound of average cooperative sensing area (UB-ACSA) is used to measure the cooperative sensing performance of the JSC UAV network \cite{Chen2019Per}, which is related to the mutual sensing interference and communication capacity. The main contributions of this paper are summarized as follows. 

\begin{itemize}
	\item[1.] We design a novel JSC antenna array and present a corresponding iterative beamforming algorithm that can flexibly form SenB and ComB using multi-layer circular array and linear array, respectively. Communication and sensing can operate simultaneously utilizing the proposed antenna array.
	
	\item[2.] We model the mutual sensing interference of the JSC UAV network, which can be used to evaluate the interference of the JSC UAV network.
	
	\item[3.] We define and formulate UB-ACSA of the JSC UAV network as the cooperative sensing performance metric, taking into consideration the average mutual sensing interference. After the formula of UB-ACSA is validated numerically, we obtain the optimal number of UAVs and power allocation ratio of sensing beam power to total available power.
\end{itemize}

The remaining parts of this paper are organized as follows. In Section \ref{sec:system-model}, we describe the cooperative JSC UAV network model, the beam sharing scheme, the design of JSC antenna, the corresponding beamforming algorithms, the JSC signal waveform, and the modeling of mutual sensing interference. 
Section \ref{sec:radarrenge} provides the closed-form expressions for minimum required signal-to-interference-and-noise ratio (SINR), the maximum sensing range of UAVs, maximum cooperative range (MCR), and UB-ACSA. Section \ref{sec:Communication} formulates the outage capacity of the JSC UAV network as the communication performance metric. Subsequently, we provide the ultimate expression for MCR based on the outage capacity of the network. In section \ref{sec:numerical-result}, the numerical simulation of the previous theoretical results is presented. Section \ref{sec:conclusion} concludes this paper. 

\textbf{Notations:} Bold uppercase letters denote matrices (i.e., $\textbf{M}$); bold lowercase letters denote column vectors (i.e., $\textbf{v}$); scalers are denoted by normal font (i.e., $\gamma$); the entries of vectors or matrices are referred to with parenthesis, for instance, the $q$th entry of vector $\textbf{v}$ is $\textbf{v}(q)$, and the entry of the matrix $\textbf{M}$ at the $m$th row and $q$th column is $\textbf{M}(m,q)$; $\textbf{I}_{Q}$ is the identity matrix with dimension $Q \times Q$. Also, matrix superscripts $\left(\cdot\right)^H$, $\left(\cdot\right)^{*}$ and $\left(\cdot\right)^T$ denote Hermitian transpose, complex conjugate and transpose, respectively. Besides, we use $\left(\cdot\right)^{-1}$ to denote inverse of matrix, $\left(\cdot\right)^{\dag}$ to denote the pseudo-inverse of the matrix, $diag\left(\textbf{v}\right)$ to denote a diagonal matrix with the entries of $\textbf{v}$ on the diagonal, $E\left( \cdot \right)$ to denote the expectation of random variable, and $\left\lfloor\cdot \right\rfloor$ to denote the floor function.
\begin{table}[]
	\caption{\label{sys_para}Key parameters and abbreviations}
	\begin{center}
		\begin{tabular}{m{2cm}<{\centering}|m{6cm}<{\centering}}
			\hline
			\hline
			{\textbf{Symbol}} & {\textbf{Description}} \\
			\hline
			
			FCUAV & The UAV acting as fusion center \\
			\hline
			SU & Subordinate UAV\\
			\hline
			MCA & Maximum cooperating area\\
			\hline
			MCR & Maximum cooperating range, ${x_Q}$\\
			\hline
			SZ & Sensing zone\\
			\hline
			ACSA & Average cooperative sensing area\\
			\hline
			UB-ACSA & The upper bound of ACSA, ${\bar S_{CSA}}(M)$\\
			\hline
			ComA & Communication subarray of JSC array\\
			\hline
			SenA & Sensing subarray of JSC array\\
			\hline
			ComB & Communication beam\\
			\hline
			SenB & Sensing beam\\
			\hline
			SPR & Sensing power ratio, ${\beta _R}$\\
			\hline
			TxA & The transmitting antenna array of SenA\\
			\hline
			RxA & The receiving antenna array of SenA\\
			\hline
			S-C pair & The pair of FCUAV and an SU that is
			 transmitting sensing data\\
			\hline
			STP & Successful transmission probability\\
			\hline
			${R_g}$ & The inner radius of MCA\\
			\hline
			${R_{max}}$ & Maximum effective sensing range\\
			\hline
			$G_p$ & OFDM symbol-based radar processing gain\\
			\hline
			$R_i$ & The distance between SU$_i$ and FCUAV\\
			\hline
			$h$ & The flying height of UAVs\\
			\hline
			\hline
		\end{tabular}
	\end{center}
\end{table}

\section{System Model}\label{sec:system-model}

\subsection{Model of Cooperative JSC UAV Network} \label{subsec:network_model}

\begin{figure}[t]
	\centering
	\includegraphics[width=0.37\textheight]{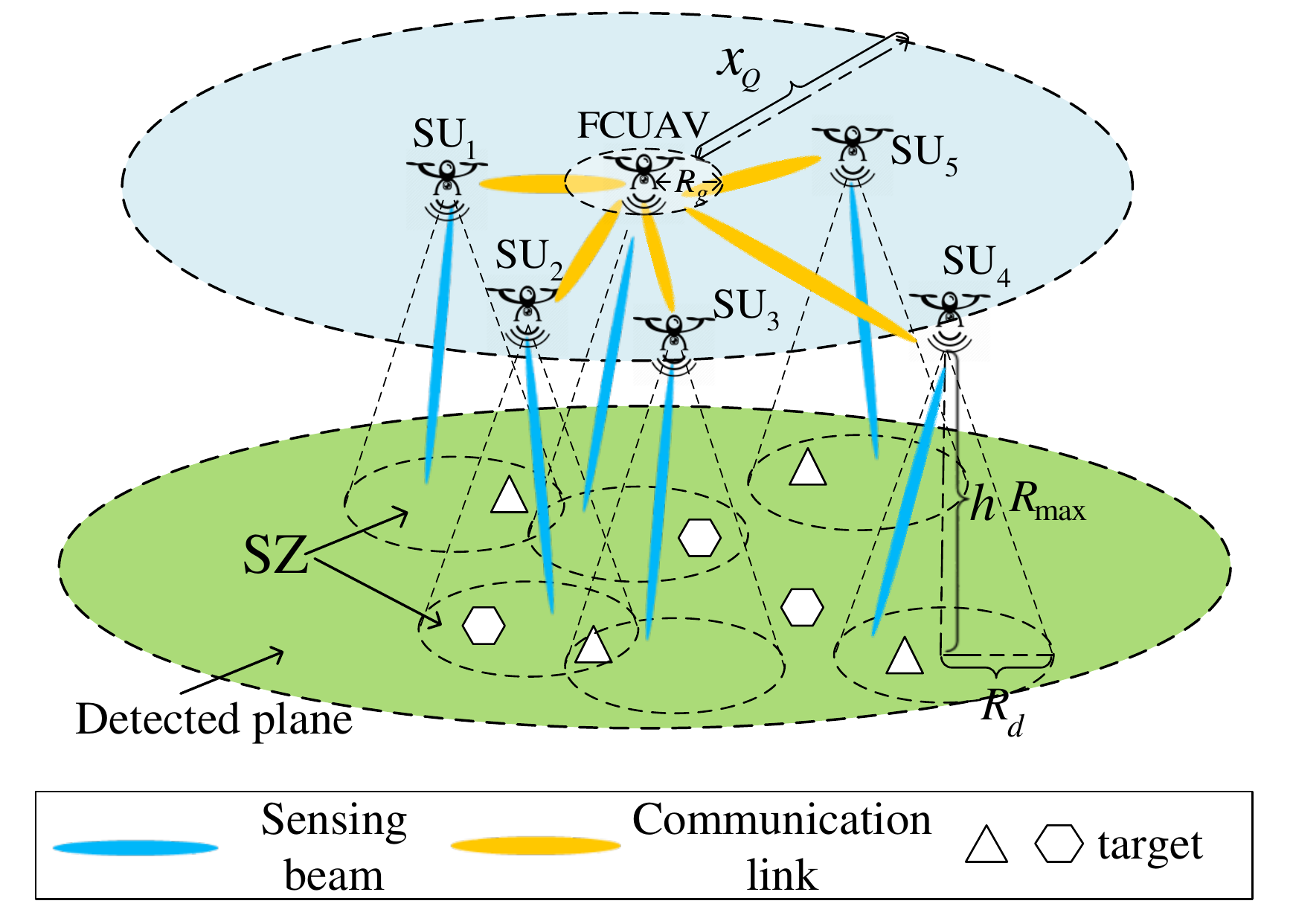}%
	\DeclareGraphicsExtensions.
	\caption{JSC UAV cooperative sensing network}
	\label{fig:Communication and radar networks}
\end{figure}

We consider a cooperative JSC UAV network that conducts downward-looking detection and detection data communication simultaneously. As illustrated in~Fig. \ref{fig:Communication and radar networks}, the network consists of an FCUAV and $M$ SUs, where all UAVs hover within a constrained two-dimensional (2D) plane at a fixed altitude $h$. Each UAV has an antenna array\footnote{All antenna elements are isotropic.} that consists of a communication array (ComA) and a sensing array (SenA) to generate a ComB and a SenB, respectively. SenB is pointed downward to detect the targets on the ground (or the sea), and ComB is pointed to UAVs to transmit sensing data. Due to the significant difference between the sensing direction and the communication direction, ComB and SenB can be easily formed orthogonally and operate simultaneously to achieve beam sharing scheme. FCUAV integrates all the sensing data from SUs with ComB and detects targets simultaneously with SenB. SUs detect the targets with SenB while simultaneously transmitting the sensing data to FCUAV with ComB. 

The antenna array of each UAV has available radio power $P$, i.e., the sum of SenB power and ComB power is $P$.
The SenB power is given by
\begin{equation}
{{\mathop{P}\nolimits} _r} = {\beta _R}P,
\end{equation}
where $\beta_R$ is defined as the sensing power ratio (SPR). The ComB power is thus $P_c = {1 - \beta _R}P$. 
OFDM waveform is adopted as the communication and sensing waveform to achieve joint communication and sensing \cite{Sturm2011Waveform}. Time division multiple access (TDMA) is adopted by the JSC UAV network to exploit the same frequency band to transmit sensing data. 
To ensure that FCUAV integrates the intact sensing data of each SU, the communication capacity between each SU and FCUAV has to be larger than the generating rate of sensing data. Therefore, the distance between FCUAV and each SU must be smaller than MCR, denoted by $x_Q$. Each UAV also has a guard radius ${R_g}$ for safety, within which no other UAVs can hover. Thus, SUs distribute within a concentric circle area centered at FCUAV, which is defined as the maximum cooperation area (MCA) whose inner and outer radii are ${R_g}$ and ${x_Q}$, respectively.

To satisfy the constraints on the false alarm rate and the detection probability for effective sensing \cite{levy2008principles}, each UAV has to decrease the loss of sensing signal power and reduce interference to other UAVs in order to maintain the minimum SINR of sensing \cite{richards2010principles}. Thus, there is the maximum sensing range for each UAV, which is denoted by $R_{max}$.
As shown in Fig. \ref{fig:Communication and radar networks}, the region on the ground that the JSC UAV network has to detect is defined as the detected plane. The region in the detected plane that can be effectively detected\footnote{i.e., the false alarm rate is below the maximum and detection probability is higher than the minimum.} by a UAV is defined as the sensing zone (SZ) of the UAV. The maximum radius of SZ is determined by $R_{max}$ and $h$. We propose the average union area of SZs of all UAVs to be the sensing performance metric of the JSC UAV network. The outage capacity of the links between SUs and FCUAV is defined as the communication performance metric of the JSC UAV network. 



{\color{blue}
}
\subsection{Beam Sharing Scheme and Design of JSC Antenna Array}\label{sec:Beam}

The beam sharing scheme is used to realize communication and sensing simultaneously. Let ($\varphi$,$\theta$) be a certain direction in the three-dimensional (3D) space, where $\theta$ denotes the elevation angle and $\varphi$ denotes the azimuth angle. ComA of each UAV generates ComB with elevation beamwidth\footnote{In this paper, the term ``beamwidth" refers to the -3dB half-power beamwidth.} $\Delta \theta_c  $ and azimuth beamwidth $\Delta \phi_c$. SenA of each UAV generates SenB with elevation beamwidth  $\Delta \theta_r$ and azimuth beamwidth $\Delta \phi_r$. Then the average directional gain of a 3D beam can be approximated by~\cite{richards2005fundamentals}~\cite{mailloux2017phased}
\begin{equation}
g \approx \frac{{26000}}{{\Delta \theta  \cdot \Delta \phi }},
\label{equ:gain_c}
\end{equation}
where $\Delta \theta$ and $\Delta \phi$ are the azimuth beamwidth and elevation beamwidth, respectively. 

\begin{figure}[t]
	\centering
	\includegraphics[width=0.38\textheight]{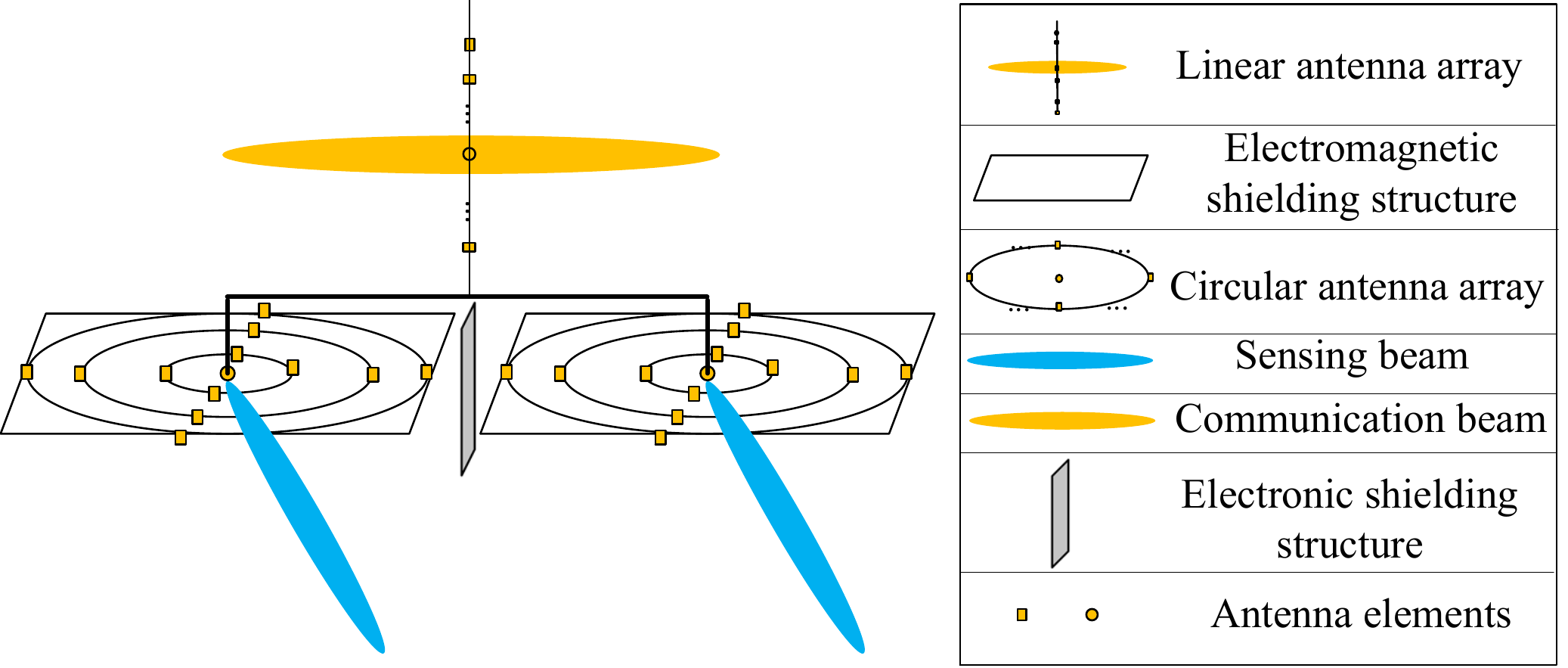}%
	\DeclareGraphicsExtensions.
	\caption{Design of JSC antenna array}
	\label{fig:JSCArray}
\end{figure}

\begin{figure}[t]
	\centering
	\includegraphics[width=0.36\textheight]{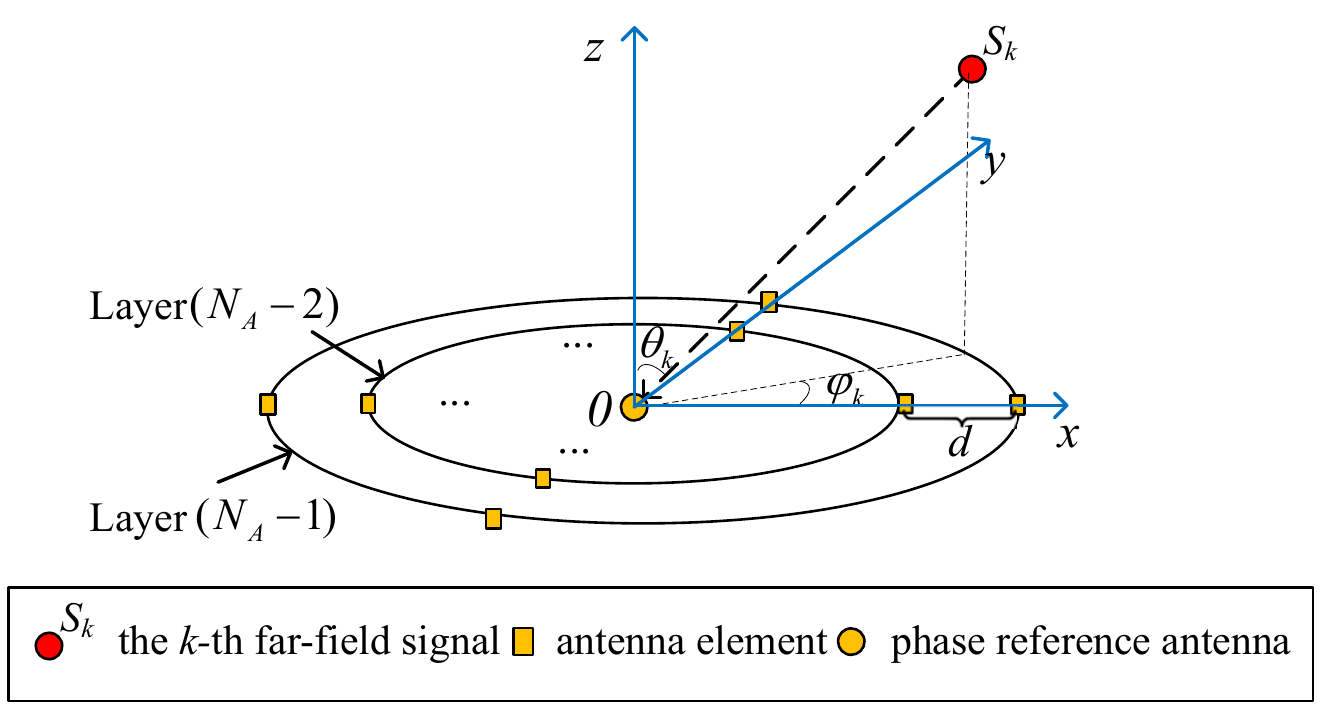}%
	\DeclareGraphicsExtensions.
	\caption{Design of JSC sensing subarray}
	\label{fig:IllustrationAntenna}
\end{figure}

\begin{figure}[t]
	\centering
	\includegraphics[width=0.22\textheight]{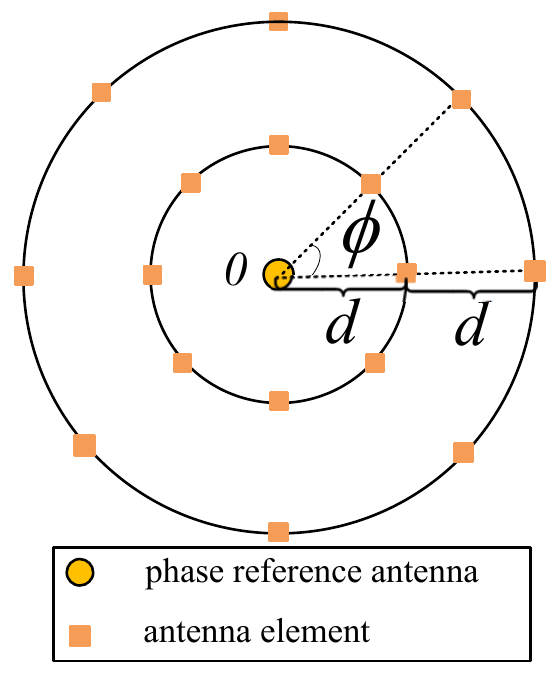}%
	\DeclareGraphicsExtensions.
	\caption{Design of adjacent layers of JSC sensing subarray}
	\label{fig:circleillustration}
\end{figure}

The antenna array designed to achieve beam sharing consists of two parts, as illustrated in Fig. \ref{fig:JSCArray}. The upper linear subarray is ComA that has $M_{com}$ antenna elements with inter-distance half-wavelength. The lower subarray is SenA composed of two decoupled circular subarrays, where one is transmitting array (TxA) and the other is receiving array (RxA). TxA and RxA are used to generate transmitting SenB and receiving SenB, respectively. The electronic shielding plate is placed between two subarrays of SenA and the pulse response between two subarrays is also accurately measured to eliminate self-interference between TxA and RxA of SenA in real-time manner~\cite{7105651}. 

Two subarrays of SenA are used to solve the problem of minimum detection range by achieving simultaneous transmitting and receiving of radar sensing signal\cite{richards2010principles}. If SenA adopts only one subarray, then it has to arrange time slots for transmitting and receiving the radar signal. In this case, the radar echo can come back to SenA before the slot for receiving, and then the missing of the target will happen. Only when the distance between UAV and the target is larger than the minimum value can the target  be possibly detected, which is inapplicable to the situation where the target is close to the JSC units. By contrast, if we adopt double subarrays to form SenA, then the radar sensing signal can be transmitted and received simultaneously. Thus, the problem of minimum detection range is solved.

As illustrated in Fig. \ref{fig:IllustrationAntenna}, SenA has antennas arranged in concentric circles. There are $N_A$ circles in each SenA, and we define each circle as a layer.  From center to periphery, SenA has the 0th layer to the ($N_A - $1)th layer. Except for the 0th layer with only one antenna element that is the phase reference antenna (PhRefA), there are ${{2}^{b}}$ antenna elements in each layer, where $b$ is a positive integer.
	
Fig. \ref{fig:circleillustration} shows the adjacent layers of SenA. The distance between the antennas that locate at the same polar angle of adjacent layers is $d$. Antennas in each layer locate at the evenly split angles, and are referred to as the $0$th to the ($2^{b}-1$)th antenna, anti-clockwise.

\subsection{Beamforming of JSC Antenna Array}\label{sec:Antenna_Beamforming}
\subsubsection{Beamforming of SenA}
Assume that there are $K_s$ planar far-field signals arriving at SenA from different directions. The angle of arrival (AoA) of the $k$th ($k=1,2,...,K_s$) signal is  $ {{\bf{p}}}_k = \left( {{\varphi }_{k}},{{\theta }_{k}} \right)^T$. We use $\Theta  = \left\{ {{{\bf{p}}_1},...,{{\bf{p}}_{K_s}}} \right\}$ to denote the set of AoAs of all $K_s$ signals. 

Let $A_{p,q}$ be the $q$th antenna in the $p$th layer of SenA. The polar angle of $A_{p,q}$ is ${\psi _{p,q}} = q \cdot \phi$, where $\phi  = \frac{{2\pi }}{{{2^b}}}$. For the $k$th signal with AoA ${{\bf{p}}}_k$, the phase difference between $A_{p,q}$ and PhRefA is given by
\begin{equation}
\begin{aligned}
{a_{p,q}}\left( {{\bf{p}}}_k \right) = \exp \left\{ { - j\frac{{2\pi }}{\lambda }{{\textbf{P} }^T_{p,q}}{{\bf v}_k}} \right\},
\label{equ:arrayphase_element}
\end{aligned}
\end{equation}
where $\lambda$ is the wavelength, and
\begin{align*}
{{\bf v} _k} &= {\left[ {\cos {\varphi _k}\sin {\theta _k},\sin {\varphi _k}\sin {\theta _k}} \right]^T},\\
{\textbf{P}_{p,q}} &= {\left[ {p \cdot d\cos \left( {\psi _{p,q}} \right),p \cdot d\sin \left( {\psi _{p,q}} \right)} \right]^T}.
\end{align*}

%
Furthermore, the steering vector of SenA is 
\begin{equation}
{\bf{a}}({{\bf{p}}_k}) = {\left[ {1,{a_{1,0}}\left( {{{\bf{p}}_k}} \right),...,{a_{{N_A} - 1,{2^b} - 1}}\left( {{{\bf{p}}_k}} \right)} \right]^T},
\label{equ:arrayphase_steering_vec}
\end{equation}
enumerating the phase differences between the PhRefA and all the other antenna elements of SenA. Then the steering matrix for $K_s$ far-field signals is ${\bf{A}} = \left[ {\bf{a}}({{\bf{p}}_1}),...,{\bf{a}}({{\bf{p}}_{K_s}}) \right]$.

The received signal vector for SenA is 
\begin{equation}
\begin{aligned}
{\bf y} = {\bf{A}}{\bf s} _r + {{\bf{n}}_s},
\label{equ:arrayphase_rawinput}
\end{aligned}
\end{equation}
where ${\bf s} _r$ is the source signal vector of dimension $K_s \times 1$, and ${{\bf{n}}_s}$ is the additive white Gaussian noise (AWGN) with covariance matrix $\sigma _r^2{{\bf{I}}_{K_s}}$ and zero mean.

Based on the least square (LS) error principle, the beamforming problem of SenA can be formulated as~\cite{shi2005new}

\begin{equation}
\begin{aligned}
\mathop {\min }\limits_{{{\bf{w}}_r}} \left\| {{{\bf{w}}^H_r}{\bf{A}} - {{\bf{r}}^{T}_d}} \right\|_2^2,
\label{equ:arrayphase_Beamforming}
\end{aligned}
\end{equation}
where ${{\bf{w}}_r}$ is the normalized beamforming vector for SenA to generate SenB, and $\textbf{r}_d$ is the desired response, i.e., 
\begin{equation}
\begin{aligned}
{{\bf{r}}_d} = diag\left( {{{\bf{r}}_{ad}}} \right) \times {{\bf{r}}_{pd}},
\label{equ:arrayphase_response}
\end{aligned}
\end{equation}
where
\begin{equation*}
{{\bf{r}}_{ad}} = \left[ {{r_{ad}}\left({{{\bf{p}}_1}} \right),...,{r_{ad}}\left( {{{\bf{p}}_k}} \right)} \right]^T,
\end{equation*}
\begin{equation*}
{{\bf{r}}_{pd}} = \left[ {{r_{pd}}\left( {{{\bf{p}}_1}} \right),...,{r_{pd}}\left( {{{\bf{p}}_k}} \right)} \right]^T,
\end{equation*}
representing the desired amplitude response and desired phase response, respectively. Note that ${{\bf{r}}_{ad}}$ is a column vector of real values, and ${{\bf{r}}_{pd}}$ is a column vector of complex values with unit modulus.

\begin{algorithm}
	\caption{Two-step Iterative LS Beamforming Algorithm}
	\label{alg:TSILSB}
	\KwIn{The desired AoAs $\{ {\bf{p}}_{1},{\bf{p}}_{2},...,{\bf{p}}_{L}\}$, and the corresponding steering matrix ${{{\bf{A}}}}$.
		\\\qquad\quad The desired amplitude pattern ${{\bf{r}}_{ad}}$.
		\\\qquad\quad Initial beamforming vector ${{\bf w}_{r,0}}$.
		\\\qquad\quad Iteration index $m = 0$.
		\\\qquad\quad Iteration time threshold $\tau_{th}$.}
	\KwOut{Final converged beamforming vector ${{\bf w}_{r,m}}$.}
	\While{${{\bf w}_{r,m}}$ {\rm does not converge and}  $m \le \tau_{th}$}{
		1) $m = m + 1$.
		
		2) ${{\bf{r}}_{pd}} = {\left[ {diag\left( {{{\bf{r}}_{ad}}} \right)} \right]^{ - 1}}{\bf{A}^ {\it T}}{ {\bf w}_{r,m - 1}^{*}}$.
		
		3) ${{\bf{r}}_{pd1}} = \frac{{{\bf{r}}_{pd}}}{|{{\bf{r}}_{pd}}|}$.
		
		4) ${{\bf w}_{r,m}}{\rm{ = }}{\left( {{{\bf{A}}^H}} \right)^\dag }{\left[ {diag\left( {{{\bf{r}}_{ad}}} \right)} \right]^{ H}}{{\bf{r}}_{pd1}^{*}}$.
	}
	\Return ${{\bf w}_{r,m}}$
\end{algorithm}

Two-step iterative LS beamforming method is presented to generate high-performance beam given AoAs~\cite{shi2005new}, as shown in \textbf{Algorithm} \ref{alg:TSILSB}.  First, we determine the desired AoAs, amplitude response ${{\bf{r}}_{ad}}$ and the corresponding steering matrix ${{{\bf{A}}}}$. Then we set the initial beamforming vector as ${\bf w}_{r,0} = \textbf{0}$ and set the iteration time threshold as $\tau_{th}$. If the iteration time $m$ is smaller than $\tau_{th}$, then the phase response ${{\bf{r}}_{pd}}$ and the beamforming vector ${\bf w}_{r,m}$ are updated iteratively. After the iteration time reaches $\tau_{th}$ or ${\bf w}_{r,m}$ converges to a stable value, ${\bf w}_{r,m}$ is the output as the beamforming vector for SenB beamforming. In \textbf{Algorithm}~\ref{alg:TSILSB}, $\left| {\bf{x}} \right|$ brings out the vector where the $i$th entry is the modulus of the $i$th entry of ${\bf{x}}$ (${\bf{x}}$ can degrade to scalar), and $\frac{{{\bf{r}}_{pd}}}{|{{\bf{r}}_{pd}}|}$ is entry-wise division.

\subsubsection{Beamforming of ComA}

Assume that there are $K_c$ planar far-field signals arriving at ComA. The $i$th signal's AoA is $\theta_i$.  Similar to SenA, the steering vector of ComA is
\begin{equation}
{\overline{\bf{a}}{({\theta _i})}} = {\left[ {1,{e^{ j\frac{{2\pi }}{\lambda }d_c\cos {\theta _i}}},...,{e^{ j\frac{{2\pi }}{\lambda }\left( {M_{com} - 1} \right) d_c \cos {\theta _i}}}} \right]^T},
\end{equation}
where $d_c$ is the distance between the adjacent antennas of ComA. Then we obtain the steering matrix of ComA as ${\overline{\bf{A}}} = \left[ {\overline{\bf{a}}{({\theta _1})}},...,{\overline{\bf{a}}{({\theta _{K_c}})}} \right]$.  Furthermore, the received signal of ComA can be expressed as ${{\bf y}_c } = {\overline{\bf{A}}} {{\textbf{s}}}_c + {{\bf n}_c}$, where ${{\textbf{s}}}_c$ is the source signal vector of dimension $K_c \times 1$, and ${{\bf{n}}_c}$ is AWGN with covariance matrix $\sigma _c^2{{\bf{I}}_{K_c}}$ and zero mean. The objective for the ComA beamforming is given by 
\begin{equation}
\begin{aligned}
\mathop {\min }\limits_{{{\bf w}_c}} \left\| {{{\bf w}^H_c}{\overline{\bf{A}}} - {{\overline{\bf{r}}}^T_{pd}} \times diag\left( {{\overline{\bf{r}}_{ad}}} \right)} \right\|_2^2,
\label{equ:Beamforming_ComA}
\end{aligned}
\end{equation}
where ${\overline{\bf{r}}_{ad}}$ and ${{\overline{\bf{r}}}_{pd}}$ are the desired amplitude response and desired phase response of ComB, respectively. Note that ${\overline{\bf{r}}_{ad}}$ is a column vector of real values, and ${{\overline{\bf{r}}}_{pd}}$ is a column vector of complex values with unit modulus. \textbf{Algorithm} \ref{alg:TSILSB} is also applied to generate ComB. After $\tau_{th}$ iterations or ${\bf w}_c$ converges to a stable value, the desired beamforming vector for ComB will be output.

\subsection{Signal Model of the JSC UAV Network} \label{sec:The Signal Model of JSC UAV Network}
OFDM signal is adopted in the JSC UAV network to exploit its advantage of the frequency diversity and obtain high processing gain~\cite{sturm2010performance}. The baseband OFDM JSC signal is modeled as \cite{sturm2010performance}
\begin{equation}
x\left( t \right) \!= \! {\sum\limits_{m = 0}^{{M_{s}}-1}} {\sum\limits_{q = 0}^{{N_c} - 1} {d_{Tx}\left( {m{N_c} \!+\! q} \right)\exp \left( {j2\pi {f_q}t} \right)}} rect\!\left(\! {\frac{{t - m{T}}}{{{T}}}} \!\right)\!,
\label{equ:OFDM_signal}
\end{equation}
where ${M_{s}}$ is the number of OFDM symbols in one frame, ${N_c}$ is the number of subcarriers, $d_{Tx}\left( {m{N_c} + q} \right)$ is the transmit symbol on the $q$th subcarrier of the $m$th OFDM block, $B$ is the bandwidth of the JSC signal, ${f_q} =  \frac{qB}{{{N_c}}}$ is the baseband frequency of the $q$th subcarrier, and $T$ is the duration time of each OFDM symbol that contains guard interval and the duration of OFDM block.

The phase difference between the transmitted and the received OFDM symbols is used to estimate the Doppler frequency shift and the time delay between the target and the UAV.

The Doppler shift results from the radial relative velocity between the UAV and the target, and can be presented as 
\begin{equation}
{f_{d,s}} = \frac{{2{v_{rel}}{f_c}}}{{{c_0}}},
\label{equ:doppler}
\end{equation}
where ${c_0}$ is the speed of light, ${f_c}$ is the carrier frequency of the OFDM signal, and ${v_{rel}}$ is the radial relative velocity between UAV and the target. The time delay is expressed as
\begin{equation}
{\tau_s} = \frac{{2R_r}}{{{c_0}}},
\label{equ:timedelay}
\end{equation}
where $R_r$ is the distance between the target and the UAV. The complex ratio of the received to the transmitted OFDM signal is expressed as~\cite{sturm2010performance}
\begin{equation} \label{equ:OFDM_Rx}
\begin{split}
	{({{\bf{D}}_{div}})_{m,q}} &=  \frac{{d_{Rx}}\left( {m{N_c} + q} \right)}{{d_{Tx}}\left( {m{N_c} + q} \right)}\\
	&= ({\bf{H}})_{m,q} \exp \left( { - j2\pi{f_q} \tau_s} \right) \exp \left( {j2\pi mT{f_{d,s}}} \right),
\end{split}
\end{equation}
where $(\cdot)_{m,q}$ is the entry of a matrix at the $m$th row and the $q$th column, ${{d_{Rx}}\left( {m{N_c} + q} \right)}$ is the received OFDM symbol corresponding to ${{d_{Tx}}\left( {m{N_c} + q} \right)}$, and  $({\bf{H}})_{m,q}$ is the complex fading factor for the $q$th subcarrier of the $m$th OFDM symbol. The matrix form of \eqref{equ:OFDM_Rx} is given by
\begin{align}
{({{\bf{D}}_{div}})_{m,q}} = ({\bf{H}})_{m,q}( {{\bf k} _D}{{\bf k}^{T}_R} )_{m,q},
\end{align}
where $m=0,1,...,M_s-1$, $q = 0,1,...,N_c-1$, ${{\bf k}_R}\left( q \right) = \exp \left( { - j2\pi {f_q} \tau_s } \right)$, and ${{{ {\bf k}_D}}}\left( m \right) = \exp \left( {j2\pi mT{f_{d,s}}} \right)$. 

By applying discrete Fourier transform (DFT) to each column of ${{\bf{D}}_{div}}$ and inverse discrete Fourier transform (IDFT) to each row of ${{\bf{D}}_{div}}$, ${f_{d,s}}$ and ${R_r}$ can be obtained respectively. 

Let ${{\bf{D}}_{m,IDFT}}$ denote the IDFT of the $m$th row of ${{\bf{D}}_{div}}$, and ${\bf{D}}_{q,DFT}$ denote the DFT of the $q$th column of ${{\bf{D}}_{div}}$. The index corresponding to the peak of ${{\bf{D}}_{m,IDFT}}$ depends on $f_{d,s}$ and can be obtained by one-dimensional exhaustive search as follows~\cite{Sturm2011Waveform}:
\begin{equation} \label{equ:ind_d,m}
in{d_{d,m}} = \left\lfloor {{f_{d,s}}T{M_s}} \right\rfloor, \ in{d_{d,m}} = 0,1,...,N_c-1.
\end{equation}

Similarly, the index corresponding to the peak of ${\bf{D}}_{q,DFT}$ depends on $R_r$ and can be obtained by one-dimensional exhaustive search as follows~\cite{Sturm2011Waveform}:
\begin{equation} \label{equ:ind_R,q}
in{d_{R,q}} = \left\lfloor {\frac{{2{R_r}B}}{{{c_0}}}} \right\rfloor, \ in{d_{R,q}} = 0,1,...,M_s-1.
\end{equation}

Based on \eqref{equ:ind_d,m} and \eqref{equ:ind_R,q}, we can obtain $f_{d,s}$ and $R_r$, respectively. The accuracy of the estimation depends on the duration time of each OFDM symbol and the bandwidth of signal \cite{Sturm2011Waveform}. This radar echo processing technique can generate a processing gain as~\cite{Sturm2011Waveform}
\begin{align}\label{equ:G_p}
G_{p}=N_{c}M_{s}.
\end{align}

\subsection{Modeling of Mutual Sensing Interference}
In order to model the mutual sensing interference, the power spectrum of the JSC OFDM signal needs to be considered first, which is the sum of the power spectra of $N_{c}$ subcarriers~\cite{proakis1994communication}. According to~\cite{proakis1994communication}, the power spectrum of the $q$th subcarrier is
\begin{align}
	{S_q}\left( f' \right) = {T_s}{\left[ {\frac{{\sin \left( {\pi \left( {f'{T_s}{N_c} - q} \right)} \right)}}{{\pi \left( {f'{T_s}{N_c} - q} \right)}}} \right]^2},
\end{align}
where $T_s = \frac{1}{B}$ is the sampling cycle of the OFDM signal. The aggregate power spectrum of all subcarriers is noise-like~\cite{Sturm2011Waveform}. For reasonable simplification, we assume that the transmitting power of the OFDM sensing signal concentrates in the baseband spectrum $[0, B]$. UAVs receive the sensing interference power imposed by all other UAVs' scattered sensing signals. 
\begin{figure}[!t]
	\centering
	\includegraphics[width=0.37\textheight]{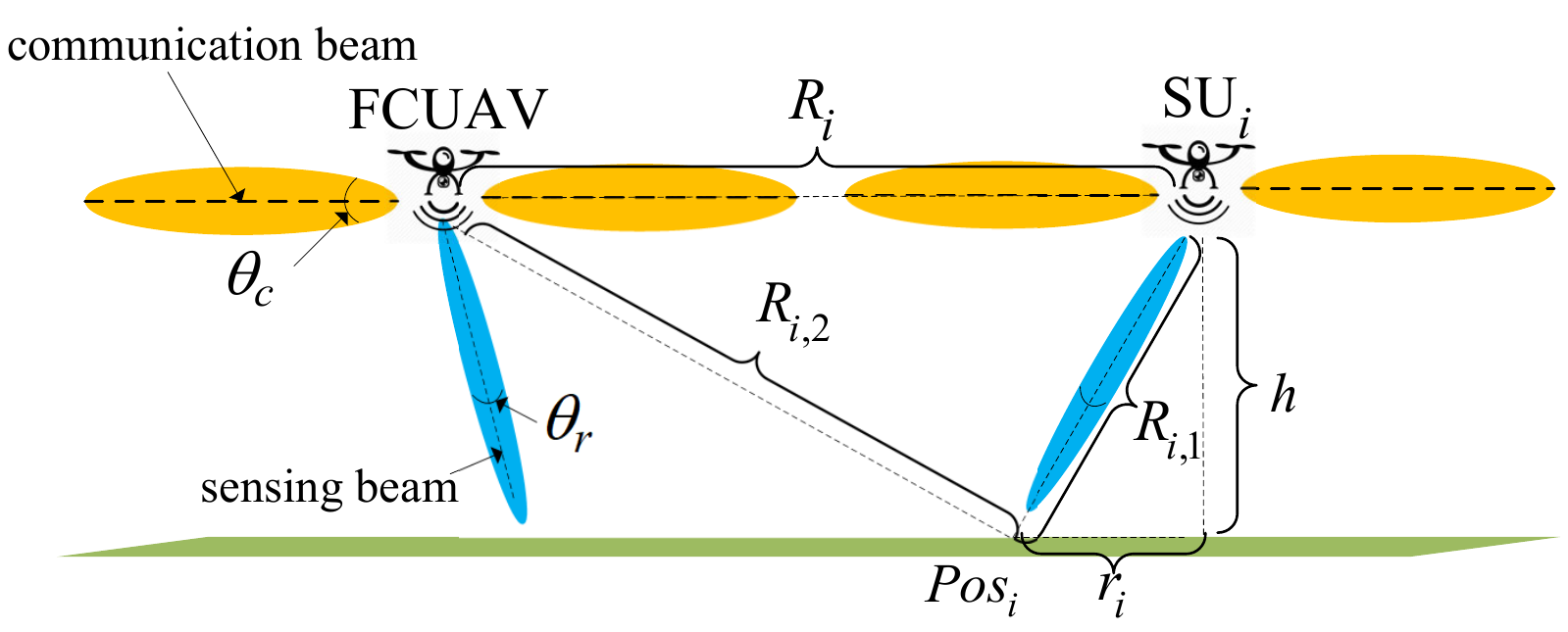}%
	\DeclareGraphicsExtensions.
	\caption{Front view of vertical pattern of sensing beam and communication beam}
	\label{fig:Directional Antenna}
\end{figure}

FCUAV is chosen as the reference to analyze the average sensing interference that a UAV receives from other UAVs. As illustrated in Fig. \ref{fig:Directional Antenna}, denote ${R_i}$ as the distance between SU$\rm _{\it i}$ and FCUAV,\footnote{$R_i$ is an independently and identically distributed (i.i.d.) variable.} $Pos_i$ as the intersection point between the SenB direction of SU$_i$ and SZ of SU$_i$, ${R_{i,1}}$ as the distance between SU$\rm _{\it i}$ and $Pos_i$, ${R_{i,2}}$ as the distance between $Pos_i$ and FCUAV, and $r_i$ as the distance between $Pos_i$ and the projection point of SU$_i$ on SZ. Assuming that the interfering sensing signals are  random due to the random scattering, the sensing interfering power imposed by SU$_i$ on FCUAV is formulated as~\cite{richards2005fundamentals,molisch2012wireless}
\begin{equation}
{I_{sen,i}} = \frac{{{P_r}{g_{ts}}}}{{4\pi }}{R_{i,1}}^{ - 2}\overline \sigma  \frac{1}{{4\pi }}{R_{i,2}}^{ - 2}\frac{{{\lambda ^2}{g_{rs}}}}{{4\pi }},
\end{equation}
where ${g_{ts}}$ and ${g_{rs}}$ are the gains of transmitting and receiving sensing beams respectively, $\overline \sigma$ is the mean of radar cross section of target, $\lambda$ is the wavelength of the carrier of JSC transceiver, and $R_{i,1}$ and $R_{i,2}$ are expressed as follows:
 \begin{equation}
\begin{array}{l}
{R_{i,1}} = \sqrt {{r_i^2} + {h^2}}, \\
{R_{i,2}} = \sqrt {{{({R_i} - r_i)}^2} + {h^2}}.
\end{array}
\end{equation}

The maximal point of $I_{sen,i}$ with regard to $r_i$ is $r_i = \frac{{{R_i}}}{2}$, which is easily obtained by the first-order and the second-order derivative of $I_{sen,i}$ versus $r_i$. Thus, we set ${R_{i,1}} = {R_{i,2}} = {\left[ {{{\left( {\frac{{{R_i}}}{2}} \right)}^2} + {h^2}} \right]^{\frac{1}{2}}}$ to obtain the upper bound of the average sensing interference imposed on FCUAV by SU$_i$.

The expectation of sensing interference imposed on FCUAV by $M$ SUs can be upper-bounded by
\begin{equation} 
\begin{split}
	{I_{sen}} &= {\sum_{i=1}^{M}E\{I_{sen,i}\}}\\
	&= M \times E\left\{ {\sum\limits_{i = 1}^M {\frac{1}{M}\frac{{{P_r}{g_{ts}}{g_{rs}}}}{{{{ {4\pi }}^3}}}{{\left( {{R_i}^2 + 4{h^2}} \right)}^{ - 2}}{\lambda ^2}\overline \sigma  } } \right\}.
\end{split}
\label{equ:Isen}
\end{equation}

The second order derivative of ${{\left( {{R_i}^2 + 4{h^2}} \right)}^{ - 2}}$ with regard to $R_i$ is $24{R_i}^2({R_i}^2+4h^2)^{-4}$ for $R_i,h>0$, which is larger than 0. Thus, ${I_{sen}}$ is convex of $R_i$ for $R_i>0$. By using the Jensen inequality, we have
\begin{equation} 
\begin{split}
{I_{sen}} &= \frac{{{P_r}{\lambda ^2}\overline \sigma  {g_{ts}}{g_{rs}}}}{{4{\pi ^3}}}M \times E\left\{ {\sum\limits_{i = 1}^M {\frac{1}{M}{{\left( {{R_i}^2 + 4{h^2}} \right)}^{ - 2}}} } \right\}\\
&\le \frac{{{P_r}{\lambda ^2}\overline \sigma  {g_{ts}}{g_{rs}}}}{{4{\pi ^3}}}\sum\limits_{i = 1}^M {E\left\{ {{{\left( {{R_i}^2 + 4{h^2}} \right)}^{^{ - 2}}}} \right\}},
\end{split}
\end{equation}
where ${E\left\{ {{{\left( {{R_i}^2 + 4{h^2}} \right)}^{^{ - 2}}}} \right\}}$ can be obtained as
\begin{align} \label{equ:ERi_a}
E\left\{ {{{\left( {{R_i}^2 + 4{h^2}} \right)}^{ - 2}}} \right\} = \frac{{{{\left( {{R_g}^2 + 4{h^2}} \right)}^{ - 1}} - {{\left( {{x_Q}^2 + 4{h^2}} \right)}^{ - 1}}}}{{{x_Q}^2 - {R_g}^2}}.
\end{align} 
The proof of \eqref{equ:ERi_a} is given in Theorem~\ref{Theo:cumulative density} in appendix. Then we have 
\begin{equation}\label{equ:I_sen}
{I_{sen}} = M\frac{{{P_r}{\lambda ^2}\overline \sigma  {g_{ts}}{g_{rs}}}}{{4{\pi ^3}}}\frac{\left[ {{{\left( {{R_g}^2 + 4{h^2}} \right)}^{ - 1}} - {{\left( {{x_Q}^2 + 4{h^2}} \right)}^{ - 1}}} \right]}{{{x_Q}^2 - {R_g}^2}}.
\end{equation}


\section{Sensing Performance Analysis of the JSC UAV Network}\label{sec:radarrenge}

\subsection{Minimum Required SINR for Effective Sensing}


As stated before, the power spectrum of OFDM signal is noise-like. Thus, the interference-and-noise imposed on sensing receiver of each UAV follows the Gaussian distribution, i.e., ${A_{in}} \sim N\left( 0,N_s + {I_{sen}} \right)$\footnote{$N\left(\mu, \sigma^2 \right)$ is the Gaussian distribution with $\mu$ as mean and $\sigma^2$ as variance.},
where $N_s$ is the thermal noise power. 

The useful sensing signal power is
\begin{equation}
S = {{G_p} {\gamma_{s}} } \left({N_s + {I_{sen}}}\right),
\label{equ:S}
\end{equation}
where ${\gamma_{s}}$ denotes SINR of sensing, and $G_p$ is presented in \eqref{equ:G_p}. We can obtain the following detection problem:
\begin{equation}
\begin{array}{l}
{H_1}:y = \sqrt S  + {A_{in}}\\
{H_0}:y = {A_{in}}
\end{array},
\end{equation}
where hypothesis $H_1$ is that there is a target in the direction of sensing beam, hypothesis $H_0$ is the opposite, and $y$ is the output signal of radar after processing ${M_s}$ OFDM symbols. The decision rule of whether there is a target in the sensing direction can be expressed as~\cite{levy2008principles}
\begin{equation}
y\mathop {{\textstyle{ >  \over  < }}}\limits_{{H_0}}^{{H_1}} \eta ',
\end{equation}
where $\eta '$ is the decision threshold. When $y>\eta'$, the sensor accepts the hypothesis $H_1$; otherwise, the sensor accepts the hypothesis $H_0$.

The false alarm rate and the detection probability are the metrics that have to be considered for detection~\cite{levy2008principles}, and are presented by
\begin{equation}
{P_F} = \int_{\eta '}^\infty  {f\left( {y|{H_0}} \right)} dy = Q\left( {\frac{{ \eta '}}{{\sqrt {N_s + {I_{sen}}} }}} \right)
\label{equ:PF}
\end{equation}
and
\begin{equation}
{P_D}{\rm{ = }}\int_{\eta '}^\infty  {f\left( {y|{H_1}} \right)dy}  = Q\left( {\frac{{{\eta ' - \sqrt S }}}{{\sqrt {N_s + {I_{sen}}} }}} \right),
\label{equ:PD}
\end{equation}
respectively, where $Q\left(\cdot\right)$ is the monotonically decreasing Q-function \cite{levy2008principles}. According to \eqref{equ:PF}, we have 
\begin{equation}
\eta' = {Q^{ - 1}}\left( {{P_F}} \right) {{\sqrt {N_s + {I_{sen}}} }},
\label{equ:eta_}
\end{equation}
where $Q^{-1}\left(\cdot\right)$ is the inverse function of Q-fucntion.
By substituting \eqref{equ:S} and \eqref{equ:eta_} into \eqref{equ:PD}, we can obtain $P_D$ as
\begin{equation}
{P_D} = Q\left[ {{Q^{ - 1}}\left( {{P_F}} \right) - \sqrt {G_p {\gamma_{s}}} } \right].
\label{equ:Q_PD}
\end{equation}

For effective detection, $P_F$ and $P_D$ must be constrained as~\cite{levy2008principles}
\begin{equation} \label{equ:constraints}
\left\{ \begin{array}{l}
{P_F} \le {\alpha _f}\\
{P_D} \ge {\alpha _D}
\end{array} \right.,
\end{equation}
where $\alpha _f$ is the maximum false alarm rate, and ${\alpha _D}$ is the minimum  detection probability. Based on \eqref{equ:Q_PD} and \eqref{equ:constraints}, the minimum SINR to meet the constraints can be expressed as
\begin{equation}
{\left( {\rm{SINR}} \right)_{\min }} = \frac{{\left[ {{Q^{ - 1}}\left( {{\alpha _f}} \right) - {Q^{ - 1}}\left( {{\alpha _D}} \right)} \right]^2}}{G_p}.
\label{equ:SINRmin}
\end{equation}

\subsection{Maximum Sensing Range} \label{sec:RadarDetectionRange}

According to \cite{richards2010principles}, the maximum sensing range of UAV is
\begin{equation}
{R_{max}}({\beta _R}) = {\left( {\frac{{{\beta _R}P{g_{ts}}{g_{rs}}{{(c_0/f)}^2}\bar \sigma {G_p}}}{{{{(4\pi )}^3}{{(\rm{SINR})}_{min}}\left( {k{T_0}{F_n}B + {I_{sen}}} \right){L_s}}}} \right)^{\frac{1}{4}}}
\label{RadEqu_2},
\end{equation}
where $\beta_{R}$ is SPR, $P$ is the total available radio power, ${g_{ts}}$ and ${g_{rs}}$ are the transmitting and the receiving antenna gains of SenB respectively, ${f}$ is the carrier frequency of transceiver, $ \overline \sigma$ is the mean radar cross section, $L_s$ is the aggregated power loss at the propagation medium, $I_{sen}$ is given in \eqref{equ:I_sen}, $k$ is the Boltzmann's constant, $F_n$ is the noise figure of the receiver, $B$ is the bandwidth of JSC transceiver, and ${T_0}=$ 290 K (in absolute temperature) is the standard temperature.

\begin{figure}[!t]
	\centering
	\includegraphics[width=0.35\textheight]{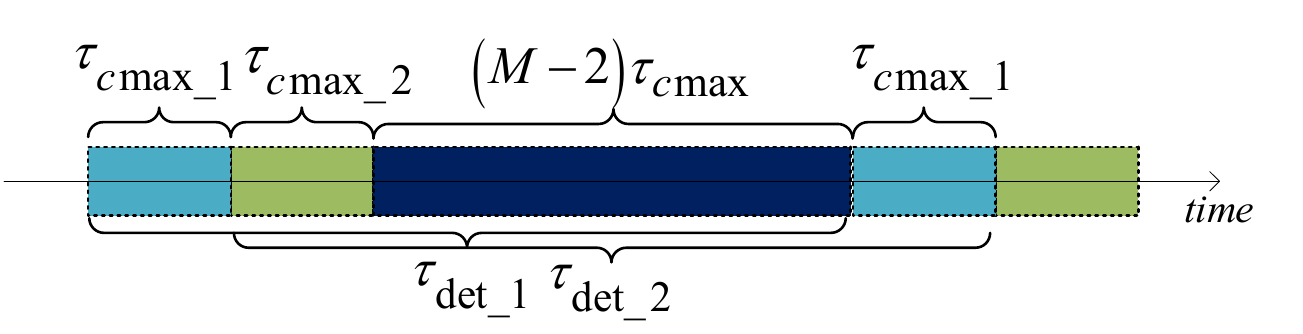}
	\DeclareGraphicsExtensions.
	\caption{Time allocation for sensing and communication}
	\label{fig_illu_tau}
\end{figure}

\subsection{Maximum Cooperative Range }
The time that SU requires to transmit the sensing data to FCUAV is defined as the ratio of the amount of sensing data to the communication throughput, which is 

\begin{equation} \label{equ:tau_c}
{\tau _c}\left( {{R_i}} \right) = \frac{{{V_{data}}{\tau _{\det }}}}{{{T_c} \left(R_i \right)}},
\end{equation}
where $V_{data}$ is the generating rate of the sensing data, $\tau_{\det }$ is the detection time, and ${{T_c} \left(R_i \right)}$ is the throughput of the communication link between FCUAV and SU$\rm _{\it i}$. With the increase of $R_i$, ${{T_c} \left(R_i \right)}$ decreases because the received ComB power decreases.

We adopt TDMA for SUs to transmit the sensing data to FCUAV. Each SU must transmit the sensing data in the assigned time slots of length ${\tau_{c\max }}$. SUs carry out downward-looking detection constantly and transmit its sensing data in the assigned slots. 

Fig.~\ref{fig_illu_tau} illustrates the time slots allocated to SUs for transmitting sensing data to FCUAV. Let ${\tau_{det\_i}}$ and ${\tau _{cmax\_i}}$ denote the sensing slots and communication slots for SU$\rm _{\it i}$, respectively. All UAVs are provided with the same sensing time ${\tau_{det}}$ and transmission time ${\tau _{cmax}}$. In order to ensure the consecutive detection of each UAV, we have
\begin{equation} \label{equ:tau_c_tau_det}
	{\tau _{cmax}} = \frac{1}{M}{\tau _{det}}.
\end{equation}
In this case, after SU$_i$ completes the sensing tasks in $\tau_{det\_i}$, the next slot for SU$_i$ to transmit sensing data comes again.

Based on \eqref{equ:tau_c}, \eqref{equ:tau_c_tau_det} and ${\tau _c}\le{\tau _{cmax}}$, we obtain ${{{{T_c} \left(R_i \right)}} \ge M \times {V_{data}}}$. Thus, the distance between FCUAV and each SU is no larger than MCR, i.e., 
\begin{equation} \label{equ:XQ_cal}
{x_Q} = {T_c}^{ - 1}\left( {M \times {V_{data}}} \right),
\end{equation}
where $T_c^{-1}(\cdot)$ is the inverse function of ${T_c}\left( R_i \right)$ and  will be presented in the next section. 

\begin{figure}[!t]
	\centering
	\includegraphics[width=0.35\textheight]{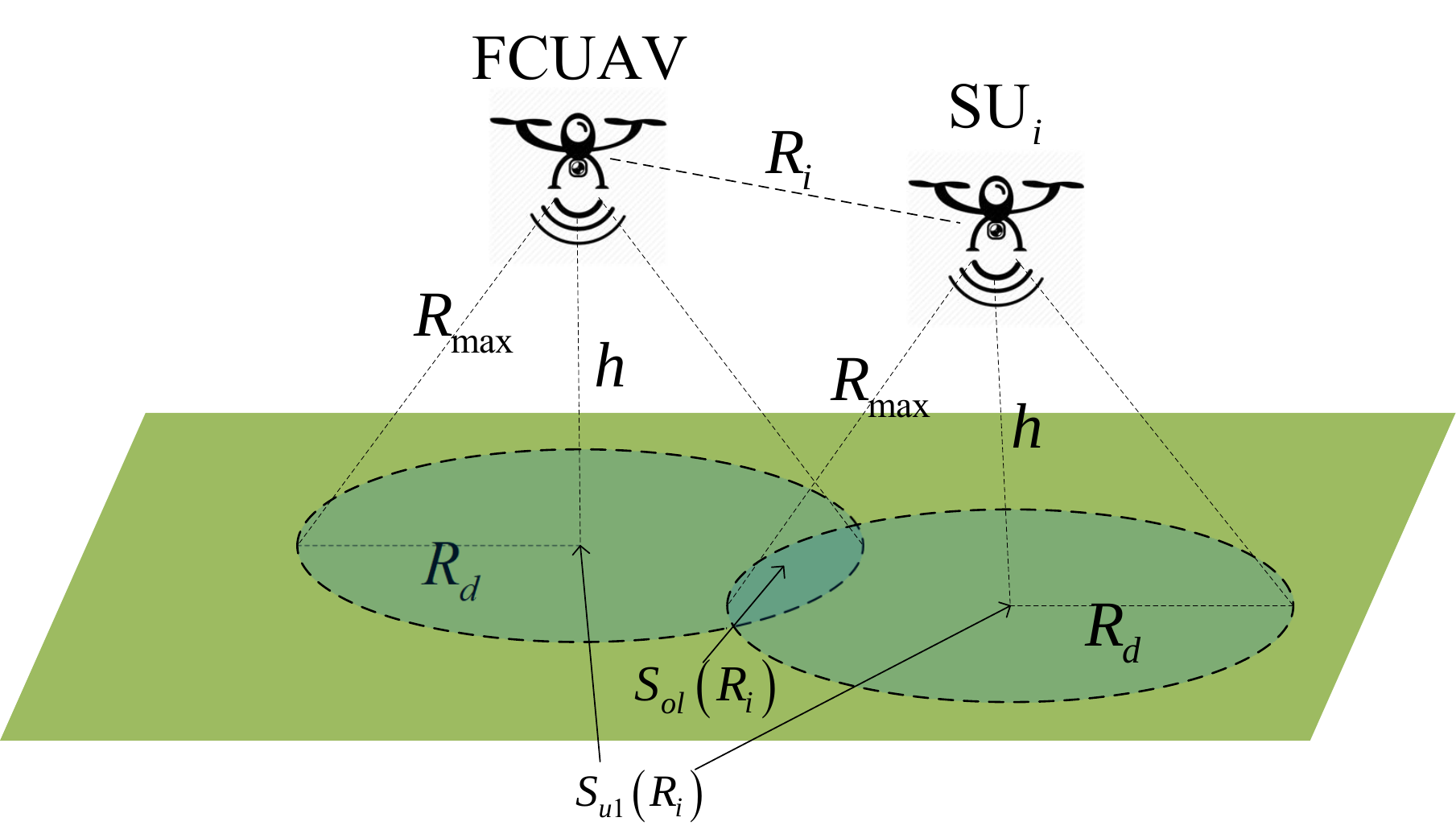}
	\DeclareGraphicsExtensions.
	\caption{The illustration of cooperative sensing of the S-C pair}
	\label{fig_Su_1}
\end{figure}

\subsection{Cooperative Sensing Performance Analysis}
We define an SU and FCUAV that are communicating sensing data as an S-C pair. As shown in Fig.~\ref{fig_Su_1}, 
the radius of SZ of each UAV is formulated as

\newcounter{mytempeqncnt}
\begin{figure*}[!ht]
	\normalsize
	\setcounter{mytempeqncnt}{\value{equation}}
	\setcounter{equation}{41} 
	\begin{equation}\label{equ:F1}
	\begin{aligned} 
	F_{1}\left({x_Q}\right) & = \frac{1}{4\left( {{x_Q}^2 - {R_g}^2} \right)}
	\Bigg\{{x_Q}\sqrt{4{R_d}^2 - {x_Q}^2} \left({2{R_d}^2 + {x_Q}^2} \right)+8\pi {R_d}^2{x_Q}^2 + 8{R_d}^2\left[{{R_g}^2\arccos\left(\frac{R_g}{2R_d}\right) - {x_Q}^2\arccos \left({\frac{x_Q}{2R_d}}\right)}\right]\\
	&\qquad\qquad+ 8{R_d}^4\left[ {2\arcsin \left( {\frac{{R_g}}{2{R_d}}} \right) + \arccot \left(\frac{\sqrt{4{R_d}^2 - {x_Q}^2}}{x_Q} \right)} -\arccot\left( {\frac{{\sqrt {4{R_d}^2 - {R_g}^2} }}{R_g}} \right)
	- 2\arcsin \left( {\frac{x_Q}{2{R_d}}} \right)\right]\\
	&\qquad\qquad- {R_g}\sqrt {4{R_d}^2 - {R_g}^2} \left( {2{R_d}^2 + {R_g}^2} \right) - 8\pi {R_d}^2{R_g}^2\Bigg\}.
	\end{aligned}
	\end{equation}
	\hrulefill 
\end{figure*}

\begin{figure*}[!ht]
	\normalsize
	\setcounter{mytempeqncnt}{\value{equation}}
	\setcounter{equation}{42} 
	\begin{equation}\label{equ:F2}
	\begin{aligned}
	{F_2}\left( {{x_Q}} \right) & = \frac{1}{{x_Q}^2 - {R_g}^2}\Bigg\{
	2\pi {R_d}^2\left( {{x_Q}^2 - {R_g}^2} \right) + \pi {R_d}^4 - 2{R_d}^4 \arccot \left( {\frac{{\sqrt{4{R_d}^2 - {R_g}^2} }}{R_g}} \right)\\
	&\qquad\qquad\qquad\quad- 2{R_d}^2\arccos \left( {\frac{{{R_g}}}{{2{R_d}}}} \right)\left( 2{R_d}^2 - {R_g}^2 \right) 
	- \frac{1}{4}{R_g}\sqrt {4{R_d}^2 - {R_g}^2} \left[ {2{R_d}^2 + {R_g}^2} \right] 
	\Bigg\}.
	\end{aligned}
	\end{equation}
	\setcounter{equation}{\value{mytempeqncnt}}
	\hrulefill 
\end{figure*}

\setcounter{mytempeqncnt}{\value{equation}}
\setcounter{equation}{37} 
\begin{equation} \label{equ:R_d}
{R_d} = \sqrt {{{({R_{\max }})}^2} - {h^2}}  \times \left[ {u\left( {{R_{\max }} - h} \right)} \right],
\end{equation}
where $u\left( \cdot \right)$ is the step function\footnote{i.e., when there is $x \ge 0$,  $u\left( x \right) = 1$; otherwise, $u\left( x \right) = 0$.}. The overlapped area between SZs of the S-C pair is 
\begin{equation}
\begin{aligned}
S_{ol}(R_i) &
= \left[ 2R_d^2 \arccos(\frac{R_i}{2R_d}) - R_i \sqrt{R_d^2 - (\frac{R_i}{2})^2 } \right] \\
&\quad\times{u\left( {2{R_d} - R_i} \right)}.
\label{S_overlap}
\end{aligned}
\end{equation}
Then, the union area of SZs of the S-C pair is
\begin{equation}
S_{u1}({R_i}) = 2\pi R_d^2 - {S_{ol}}\left( {{R_i}} \right)
\label{equ:unionArea S-C pair}.
\end{equation}

The expectation of ${S_{u1}}\left( {{R_i}} \right)$, for $ {R_g}\le R_i < x_Q$, is defined as the sensing performance metric of the S-C pair, which is formulated as \cite{Chen2019Per}
\begin{equation}\label{equ:expectation of Su1}
\begin{aligned}
{E_S}({x_Q}) &= \int_{R_g}^{{x_Q}} {{f_{R_i}}\left( r \right)S_{u1}({r}) dr}\\
&=\begin{cases}
{{F_1}\left( {{x_Q}} \right)},&{x_Q}{\rm{ < 2}}{{\rm{R}}_d}{\rm{ }}\\ 
\\
{F_2}\left( {{x_Q}} \right),&{x_Q} \ge 2{R_d}
\end{cases},
\end{aligned}
\end{equation}
where ${f_{R_i}}\left( r \right) =\frac{{2r}}{{{x_Q}^2 - {R_g}^2}}$ for ${R_g} < r < {x_Q}$, is the probability density function (PDF) of $R_i$. Moreover, $F_1\left( {{x_Q}} \right)$ and $F_2\left( {{x_Q}} \right)$ are given in \eqref{equ:F1} and \eqref{equ:F2}, respectively.

There are $M$ S-C pairs located in MCA. ACSA of the JSC UAV network is defined as the average union area of SZs of $M$ S-C pairs. Because it is extremely intractable to obtain an explicit expression of the average union area of a certain number of randomly distributed circle areas. For tractability, we merely subtract the overlapped sensing area between each SU and FCUAV to obtain the upper bound of ACSA (UB-ACSA) as the sensing performance metric of the JSC UAV network. UB-ACSA is obtained as \cite{Chen2019Per}
\setcounter{mytempeqncnt}{\value{equation}}
\setcounter{equation}{43} 
\begin{equation} \label{equ:upperboundofACSA}
\begin{split}
\overline {{S_u}} (M) &= {E_{\boldsymbol{R}}}\left\{ {\sum\limits_{i = 1}^M {\left[ { {{S_{u1}(R_i)}}  - {S\left( {{R_d}} \right)}} \right]} } \right\} + {S\left( {{R_d}} \right)}\\
&= M \times {E_S}({x_Q}) - (M - 1){S\left( {{R_d}} \right)},
\end{split}
\end{equation}
where ${S\left( {{R_d}} \right)} = \pi R_d^2$ is the area of SZ of each UAV, and $\boldsymbol{R}=$ $(R_1, R_2, ..., R_M)$ is the multiple random variables composed of the distances between SUs and FCUAV. Each element of $\boldsymbol{R}$ is an i.i.d. variable.

UB-ACSA can be achieved when all SUs do not have overlapped SZs, which requires FCUAV to coordinate the trajectory of each SU. The larger UB-ACSA is, the better the sensing performance of the JSC UAV network is. The JSC UAV network can detect the zone with area of UB-ACSA in much shorter time than individual UAV can do. The overlapped SZs of SUs can be detected by different SUs from different directions, which can reduce the probability of false detection in the overlapped SZs.

\section{Outage Capacity of the JSC UAV Network} \label{sec:Communication}

In this section, we obtain the outage capacity as the metric of communication performance, as well as the ultimate expression of $x_Q$ with respect to the number of UAVs and SPR.
\subsection{Communication Channel Model}

The power of communication signal received by FCUAV is 
\begin{equation}
{P_0} = {P_c}{g_c}{h_c}{ {{x_0}} ^{ - \alpha }},
\end{equation}
where $P_c$ is the power of ComB, $\alpha$ is the path loss exponent, ${g_c}$ is the directional gain of the communication link, $h_c$ is the small scale fading factor that follows Rician distribution with Rician factor $K_h$, and $x_0$ is the distance between the transmitter and FCUAV. We have ${g_c}{\rm{ = }}{g_{tc}} \times {g_{rc}}$, where ${g_{tc}}$ and ${g_{rc}}$ are the directional gains of transmitting ComB and receiving ComB, respectively. The PDF of $h_c$ is expressed as follows~\cite{Yuan8345703}:
\begin{equation}
{f_{{h_c}}}\left( w \right) = \frac{{\left( {{K_h} + 1} \right){e^{ - {K_h}}}}}{{\bar \Omega }}{e^{ - \frac{{\left( {{K_h}+ 1} \right)w}}{{\bar \Omega }}}}{I_0}\left( {2\sqrt {\frac{{{K_h}\left( {{K_h} + 1} \right)w}}{{\bar \Omega }}} } \right),
\end{equation}
where ${I_0}\left(x\right) = \sum_{n=0}^{\infty}{\frac{\left(x/2\right)^{2n}}{n!\Gamma(n+1)}}$ is the 0-th order modified Bessel function of the first kind, ${K_{h}} = {{{v^2}} \mathord{\left/	{\vphantom {{{v^2}} {{\sigma _K}^2}}} \right.
\kern-\nulldelimiterspace} {{\sigma _K}^2}}$, $\overline \Omega  {\rm{ = }}2{{{\sigma _K}^2}}{\rm{ + }}{v^2}$ is the normalized power, ${v^2}$ denotes the strong line-of-sight (LOS) power, and ${{\sigma _K}^2}$ represents the multipath reflected power \cite{azari2018ultra,goddemeier2015investigation,khuwaja2018survey}. 

\subsection{Successful Transmission Probability and Outage Capacity}

The successful transmission probability (STP) is defined as the probability that the received SNR (or SINR) is larger than a threshold $\gamma$ that is necessary for successful transmission \cite{Yuan8345703},\cite{Zhalehpour2015Outage}, i.e., 
\begin{equation}
\begin{split}
\rho _c^s\left( x_0, \gamma \right) &= \Pr \left( {\frac{P_0}{N_{com}} > \gamma } \right)\\
&= \Pr \left( {h_c > \frac{{\gamma N_{com}}}{{{P_c}{g_c}}}{{ {{x_0}} }^\alpha }} \right),
\end{split}
\label{equ:SuccessfulTransmission}
\end{equation}
where $N_{com}$ is AWGN of communication receiver. Since $h_c$ follows Rician distribution with Rician factor $K_h$, STP is formulated as \cite{azari2018ultra}
\begin{equation}
\rho _c^s\left( {{x_0},\gamma } \right) = Q\left( {\sqrt {2{K_h}} ,\sqrt {\frac{{2\gamma \left( {1 + {K_h}} \right){x_0}^\alpha N_{com}}}{{{g_c}{P_c}}}} } \right),
\label{equ:SuccessfulTransmissionResult}
\end{equation}
where $Q\left( {{a_1},{a_2}} \right)$ is the first order Marcum Q-function. The outage probability is the complement of $\rho _c^s$, i.e.,
\begin{equation}
\rho _c^{out} = 1 - \rho _c^s\left( {{x_0},\gamma } \right)
\label{equ:PSOUT_relation}.
\end{equation}

The outage capacity is written as~\cite{Zhalehpour2015Outage}
\begin{equation}
{T_C}\left( {{x_0}} \right) = {B}(1 - \varepsilon )\log \left( {1 + {\gamma _{th }}} \right),
\label{equ:OutThrou}
\end{equation}
where $B$ is the available bandwidth of the JSC transceiver, $\varepsilon $ is the maximum of $\rho _c^{out}$, and ${\gamma _{th }}$ is the SNR threshold that makes the outage probability equal to $\varepsilon $\cite{Zhalehpour2015Outage}. If $\gamma > \gamma_{th }$, then the outage probability will be larger than $\varepsilon$. Combining \eqref{equ:SuccessfulTransmissionResult}, \eqref{equ:PSOUT_relation} and \eqref{equ:OutThrou}, we have 
\begin{equation}
{T_c}\left( {{x_0}} \right) = B(1 - \varepsilon )\log \left( {1 + \frac{{{g_c}{P_c}{{\left[ {{Q^{ - 1}}\left( {\sqrt {2{K_h}} ,1 - \varepsilon } \right)} \right]}^2}}}{{2\left( {1 + {K_h}} \right){x_0}^\alpha N_{com}}}} \right),
\label{equ:OutThrouResult}
\end{equation}
where $Q^{-1}(\cdot,\cdot)$ is the inverse function of $Q(a_1,a_2)$ with regard to $a_2$,\footnote{As $Q\left( {{a_1},{a_2}} \right)$ is a monotonically decreasing function of ${a_2}$ \cite{Shnidman32133}, the inverse function of the first order Marcum Q-function with regard to $a_2$ exists. If $b_1 = {{Q^{ - 1}}\left( {\sqrt {2{K_h}} ,1 - \varepsilon } \right)}$, then $Q\left({\sqrt {2{K_h}}}, b_1\right) = 1-\varepsilon$.}  and  ${T_c}\left( {{x_0}} \right)$ is the communication performance metric of the JSC UAV network.

According to \eqref{equ:XQ_cal} and \eqref{equ:OutThrouResult}, we can obtain $x_Q$ as 
\begin{equation}
{x_Q} = {\left[ {\frac{{{g_c}(1 - {\beta _R})P{{\left[ {{Q^{ - 1}}\left( {\sqrt {2{K_h}} ,1 - \varepsilon } \right)} \right]}^2}}}{{2\left( {1 + {K_h}} \right)N_{com}\left( {{2^{\frac{{M{V_{data}}}}{{B\left( {1 - \varepsilon } \right)}}}} - 1} \right)}}} \right]^{1/\alpha }}
\label{equ:XQ_result}.
\end{equation}

With  \eqref{equ:expectation of Su1}, \eqref{equ:F1}, \eqref{equ:F2}, \eqref{equ:upperboundofACSA} and \eqref{equ:XQ_result}, UB-ACSA of the JSC UAV network can be calculated as the function of $M$ and $\beta_R$.
 
\section{Numerical Results}\label{sec:numerical-result}


In this section, numerical and Monte-Carlo simulations are conducted to verify the theoretical analysis in the previous sections and show the impact of the number of SUs and SPR on  UB-ACSA. The parameters used in the simulations are listed in TABLE \ref{Parameter} \cite{richards2010principles},\cite{khuwaja2018survey},\cite{lakshmanan2007overview}.

\renewcommand\arraystretch{1.05}
\begin{table}[]
	\centering
	\caption{Simulation Parameters}
	\begin{tabular}{m{2cm}<{\centering}|m{1.7cm}<{\centering}|m{3.5cm}<{\centering}}
		\hline
		\hline
		\label{Parameter:simulation}
		{\textbf{Parameter Items}}  & {\textbf{Value}} & {\textbf{Description}}\\
		\hline
		$P$ & $10$ W & The total available power\\ 
		\hline
		$N_{com}$ & $-$ 94 dB & The power of AWGN for communication\\
		\hline
		$\gamma$ &  2  & The threshold for STP\\
		\hline
		${g_{tc}}$ & 8 & The gain of transmitting ComB\\
		\hline
		${g_{rc}}$& 8 & The gain of receiving ComB\\ 
		\hline
		${M_s}$ & 16 & The number of OFDM symbols in a frame\\ 
		\hline
		${N_c} $ & 64 & The number of subcarriers\\ 
		\hline
		${G_p}$ & 1024 & The processing gain of sensing\\ 
		\hline
		${M_{com}}$ & 16 & The number of antennas of ComA\\ 
		\hline
		${N_A}$ & 17 & The number of layers of SenA\\ 
		\hline
		${2^b}$ & 16 & The number of antennas in each layer of SenA\\
		\hline
		${\alpha _f}$ & ${10}^{-8}$ & The maximum false alarm rate\\ 
		\hline
		${\alpha _D}$ & 0.99999999 & The minimum detection probability \\ 
		\hline
		$h$ & 150 m & The flying height of UAVs\\ 
		\hline
		$\alpha$ & 2.6 & The path loss exponent\\ 
		\hline
		 $\varepsilon$ & 0.1 & The maximum outage probability\\ 
		\hline
		 $f_c$& 24 GHz & The carrier frequency\\ 
		\hline
		 $\overline \sigma$ & 1 & The average radar cross section\\ 
		\hline
		 $F_n$ & 10 & The noise figure of receiver\\ 
		\hline
		 $T_0$ & 290 K & The standard temperature\\ 
		\hline
		 $B$ & 100 MHz & The bandwidth of transceiver\\ 
		\hline
		 $V_{data}$ & 1 MB/s & The generating rate of sensing data\\ 
		\hline
		 $K_h$ & 10 dB & The Rician factor \\ 
		\hline
		 $L_s$ &  1 & The propagation loss \\ 
		\hline
		\hline
	\end{tabular}
	\label{Parameter}
\end{table}

\begin{figure}[!t]
	\centering
	\includegraphics[width=0.37\textheight]{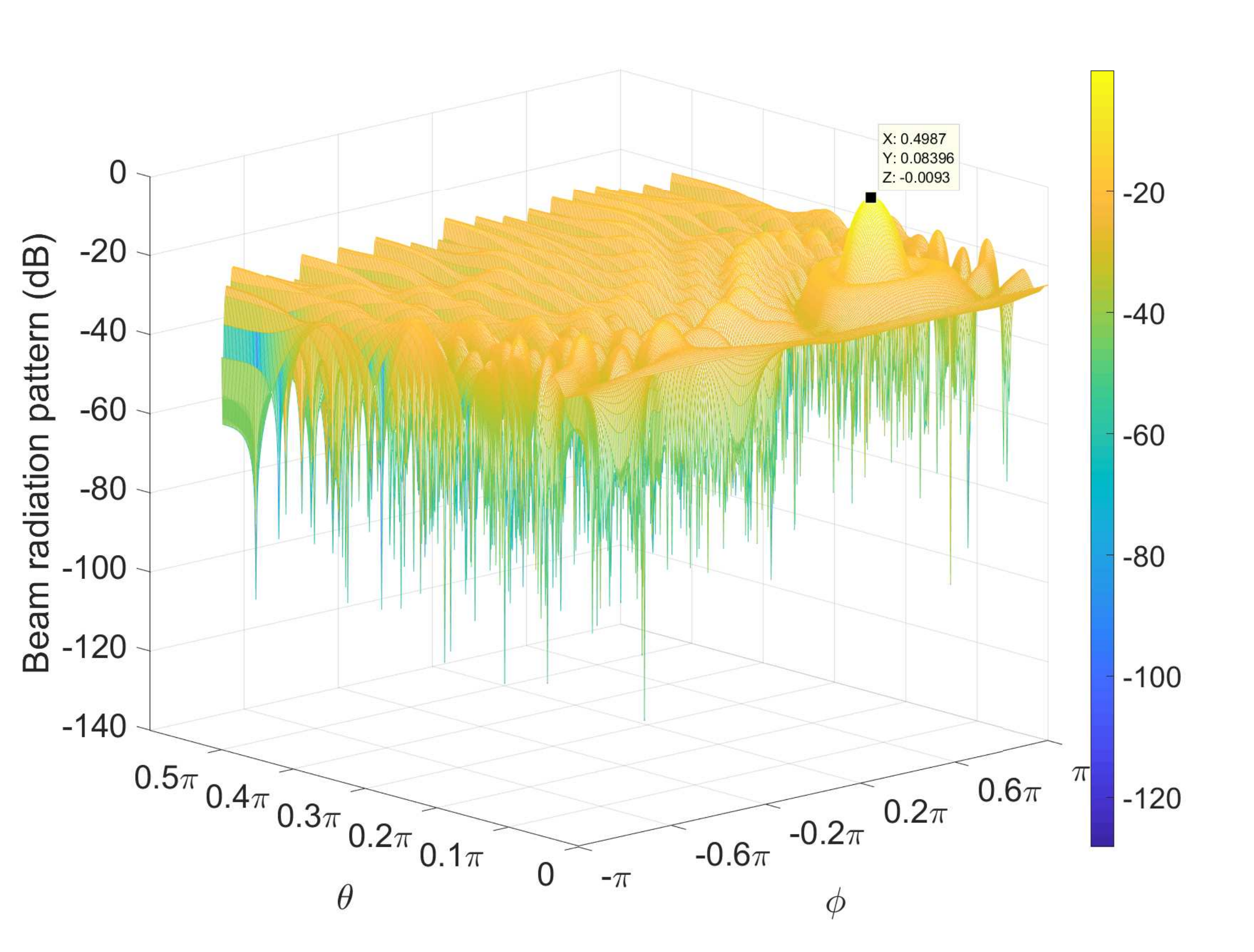}
	\DeclareGraphicsExtensions.
	\caption{3D beam radiation pattern of SenB over AoAs in radians}
	\label{fig:ArrayAntennaFig}
\end{figure}

Fig. \ref{fig:ArrayAntennaFig} shows the normalized 3D beam pattern of SenB. The mainlobe gain of 3D SenB is at least 21dB higher than that of the sidelobes. The azimuth and elevation beamwidth of SenB are approximately 4.5 and 5 degrees, respectively. Therefore, the average gain of SenB is approximately 128 (in decimal), based on \eqref{equ:gain_c}. 

\begin{figure}[!t]
	\centering
	\includegraphics[width=0.37\textheight]{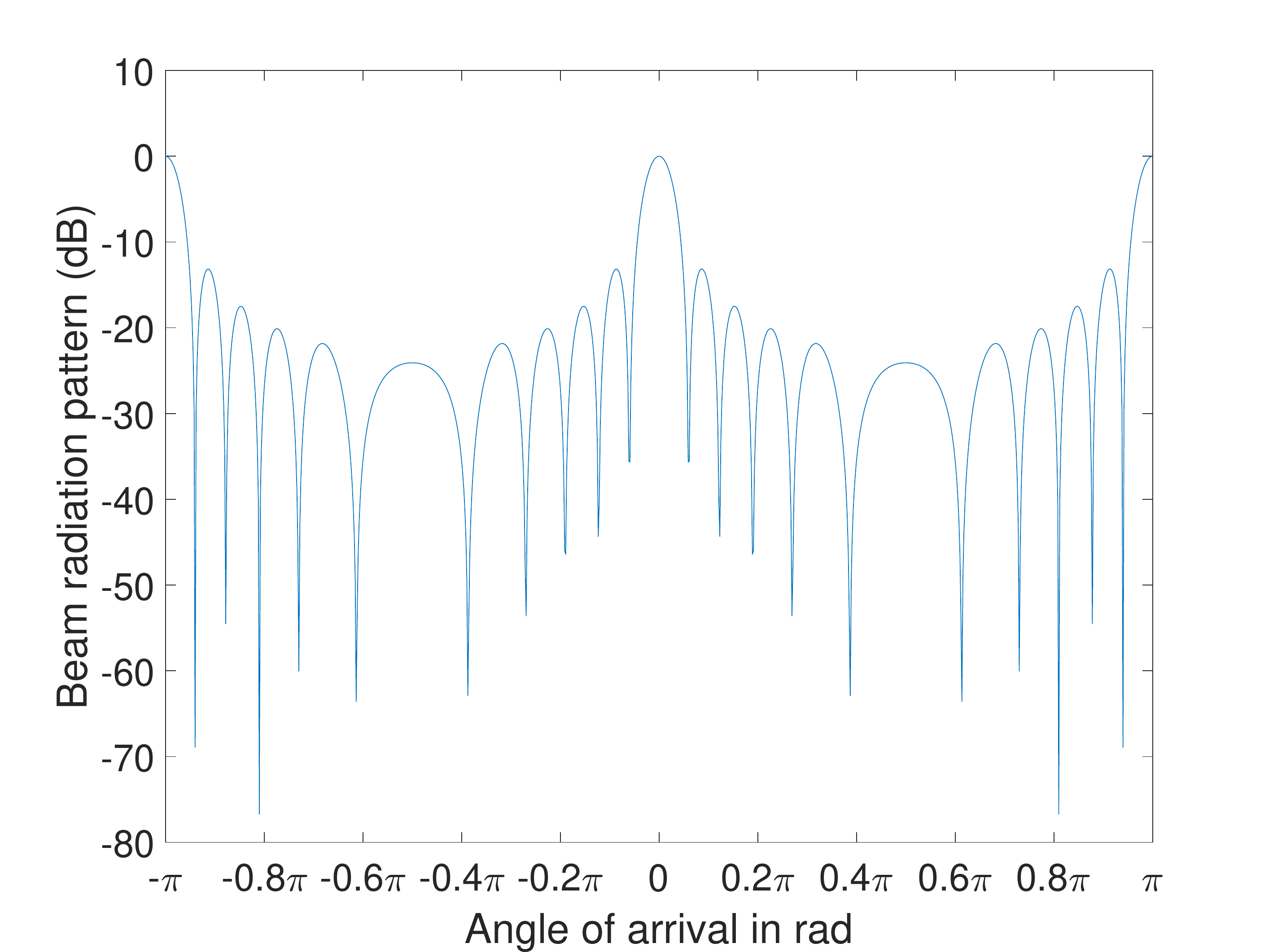}
	\DeclareGraphicsExtensions.
	\caption{Beam radiation pattern of ComB over AoAs in radians}
	\label{fig:CommAntennaFig}
\end{figure}

Fig. \ref{fig:CommAntennaFig} shows the normalized 2D beam pattern of ComA.  The mainlobe gain of ComB is at least 15 dB higher than that of the sidelobes. The elevation beamwidth of ComB is approximately 9 degrees. Besides, the azimuth beamwidth of ComB is 360 degrees. Thus, the average gain of ComB is approximately 8 (in decimal) according to \eqref{equ:gain_c}.

\begin{figure}[!t]
	\centering
	\includegraphics[width=0.38\textheight]{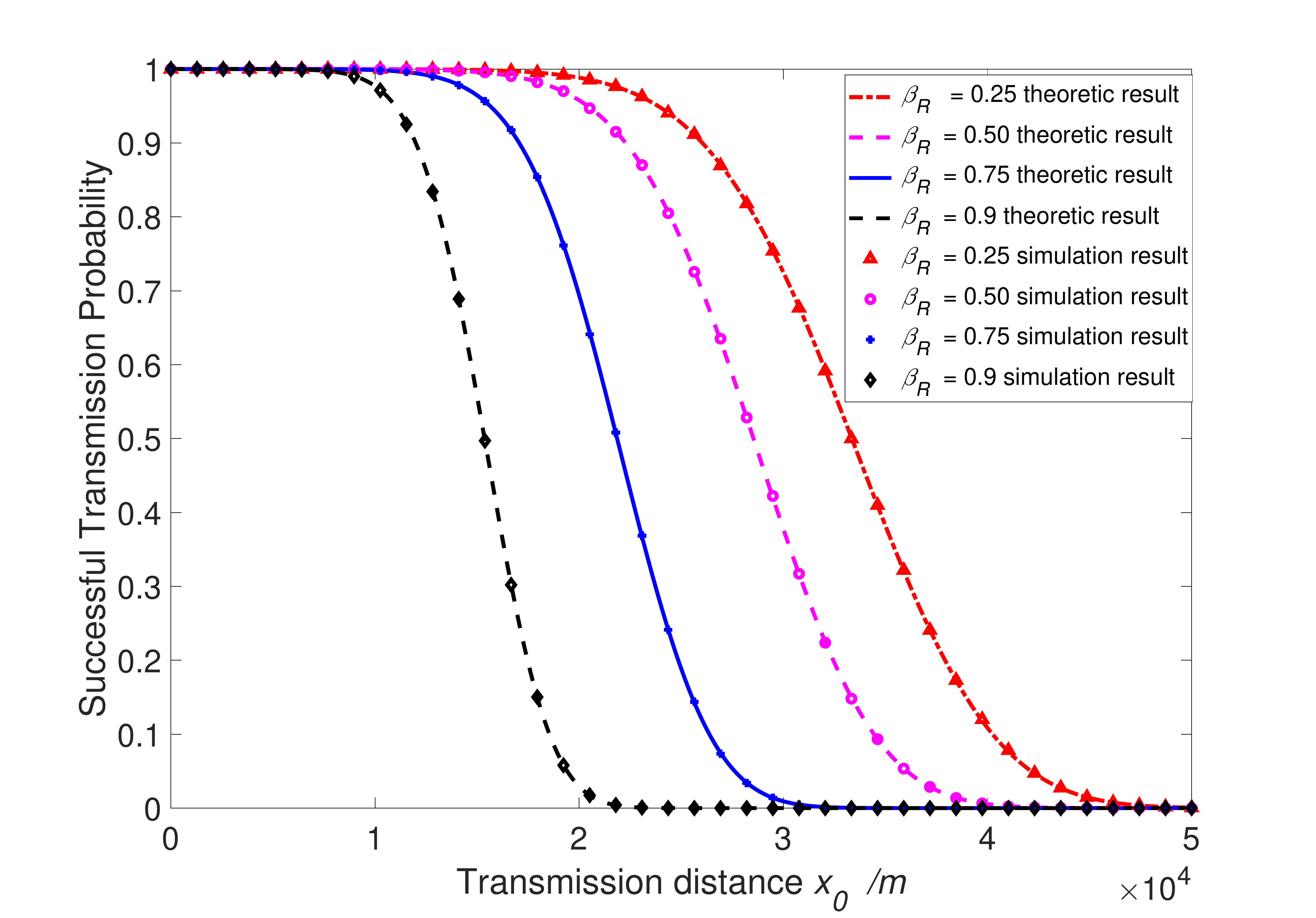}
	\DeclareGraphicsExtensions.
	\caption{STP decreases with transmission
		distance increasing under different SPRs when ${K_{h}} =$ 10}
	\label{fig:linkSuccessProb}
\end{figure}

Fig. \ref{fig:linkSuccessProb} plots both analytical and simulation results of STP versus the transmission distance under different SPR. We can see that STP is a monotonically decreasing function of the transmission distance. We also see the transmission distance becomes smaller with the increase of SPR under the same STP constraint, because the ComB power decreases as SPR or transmission distance increases. 

\begin{figure}[!t]
	\centering
	\includegraphics[width=0.38\textheight]{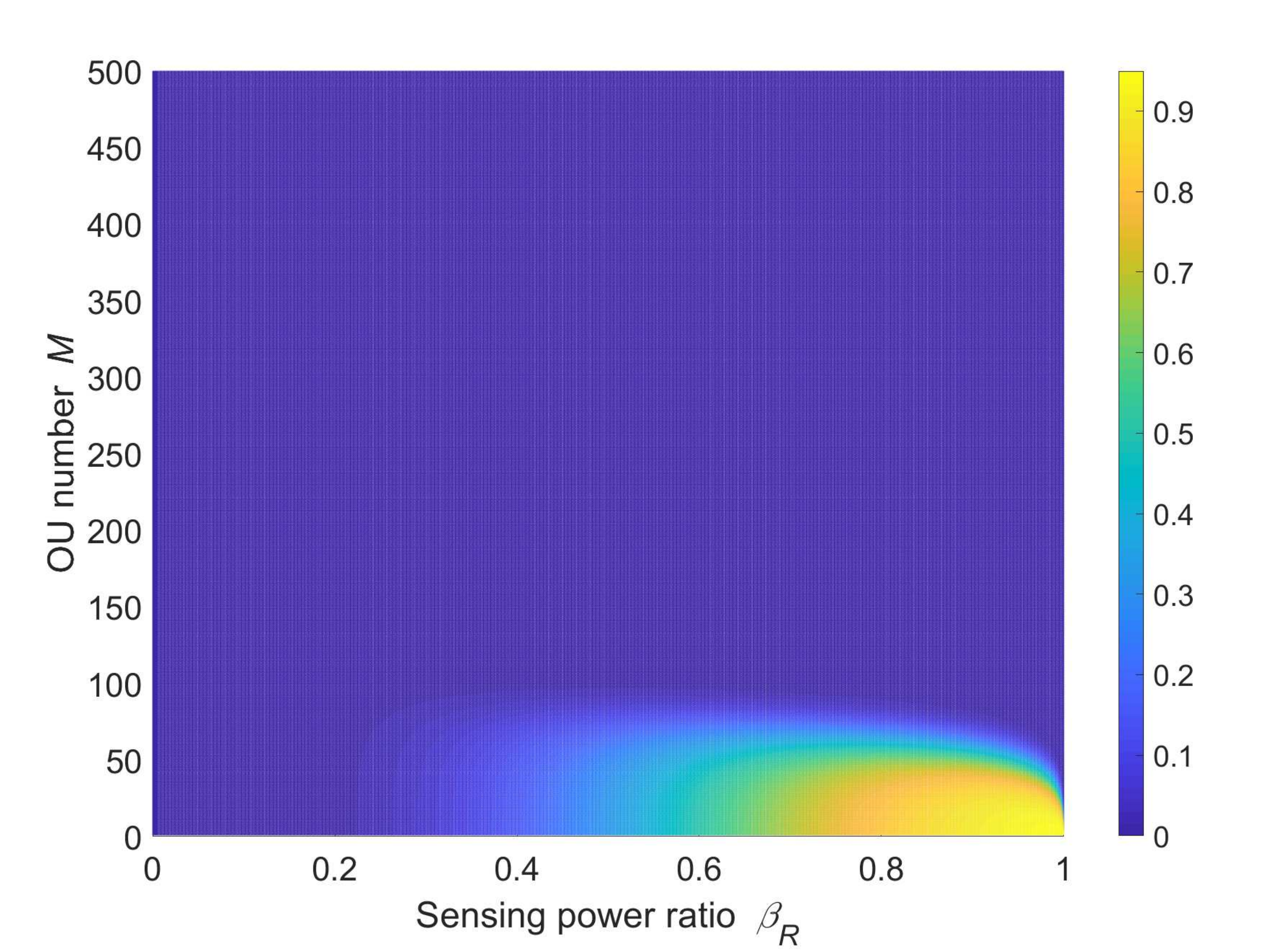}
	\DeclareGraphicsExtensions.
	\caption{$P_D$ decreases with the increase of SPR and SU number when ${K_{h}} =$ 10, $R_{max} =$  500 m, and $\alpha_f =$ 10$^{-8}$}
	\label{fig:PD_Pf}
\end{figure}

Fig. \ref{fig:PD_Pf} presents the results of $P_D$ versus $\beta_R$ and $M$. The target is at the distance of $R_{max} = $ 500 m, and the maximum false alarm rate is $\alpha_f =$ 10$^{-8}$. As is shown in Fig. \ref{fig:PD_Pf}, on the one hand, $P_D$ decreases sharply when $M$ increases to a certain value under given $\beta_R$. This is because the increase of $M$ results in the decrease of $x_Q$ and the increase of $I_{sen}$. On the other hand, when $M$ is given, $P_D$ first increases with the increase of $\beta_R$. Then after the maximum point, $P_D$ decreases. This is because as $\beta_R$ increases, the power of SenB increases at first, increasing the SINR of sensing. After $\beta_R$ gets too large, the ComB power decreases, leading to the decrease of $x_Q$ and increase of $I_{sen}$. Thus, when SINR of sensing is lower than $(\rm{SINR})_{min}$,  $P_D$ will decrease sharply with the deterioration of SINR. The decrease of $R_{max}$ is required to increase SINR of sensing when  $\beta_R$ or $M$ increases.


%
%

\begin{figure}[!t]
	\centering
	\includegraphics[width=0.38\textheight]{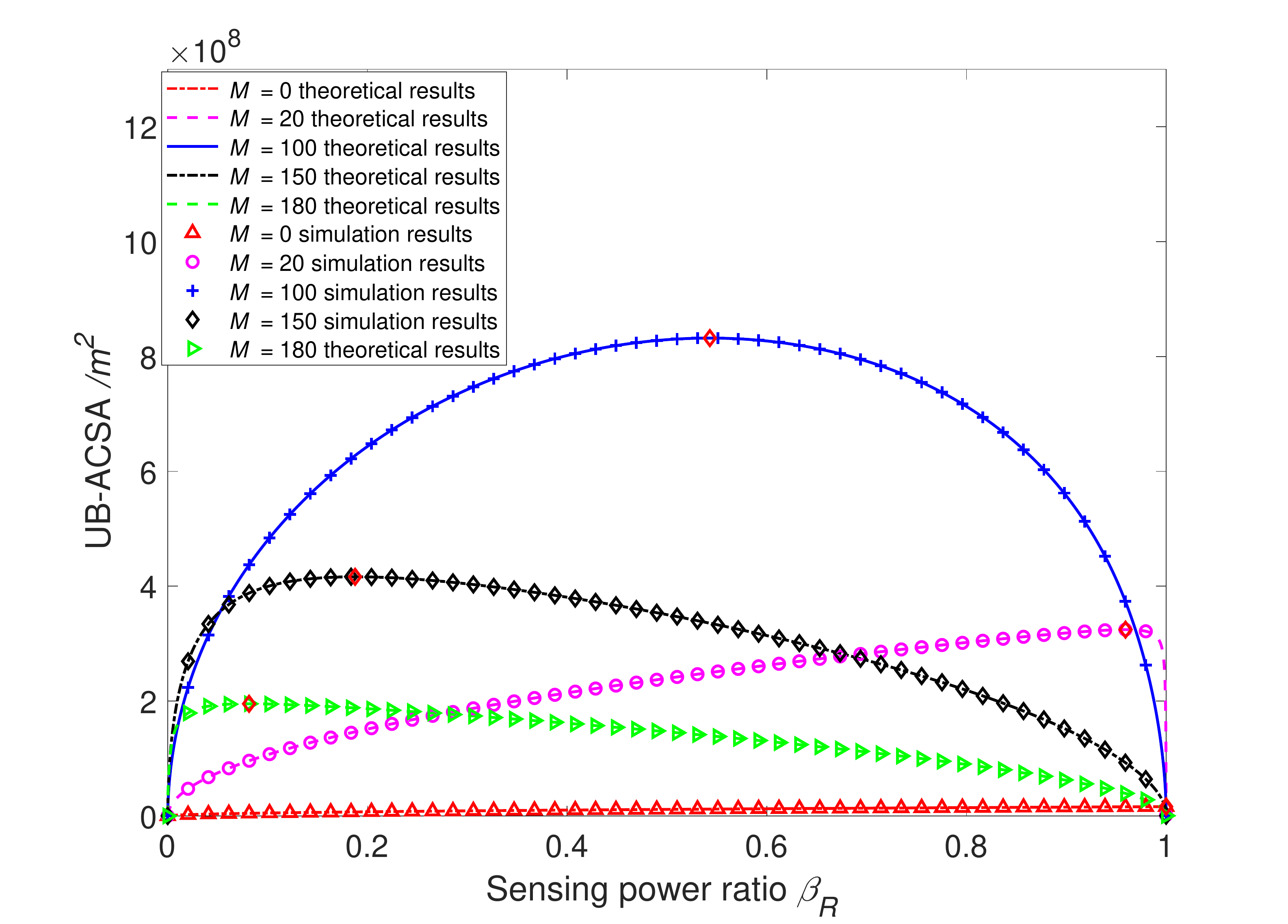}
	\DeclareGraphicsExtensions.
	\caption{UB-ACSA changes against SPR $\beta_R$ under different SU number $M$ when ${K_{h}} =$ 10}
	\label{fig:SCDAvsbeta}
\end{figure}

Fig. \ref{fig:SCDAvsbeta} plots both the analytical and simulation results of UB-ACSA changing versus  $\beta_R$, under different given $M$. Each curve can be generally divided into three stages given $M$. In the first stage where $\beta_R$ is low, the power of ComB is large. Then MCA is large enough for UAVs to expand the sensing zone. Thus, UB-ACSA increases with the increase of $\beta_R$. When $\beta_R$ is too large, the second stage comes. In the second stage, after the maximum point, UB-ACSA decreases with the increase of $\beta_R$, because the mutual sensing interference (i.e., $I_{sen}$) will increase with the decrease of $x_Q$ and the increase of the SenB power. Therefore, the maximum sensing range (i.e., $R_{max}$) decreases with the increase of $\beta_R$ in the second stage. When $\beta_R$ is so large that $R_{max} < h$ holds, the third stage appears. In the third stage, UB-ACSA decreases and converges to 0 m$ ^2$, because $I_{sen}$ is extremely large. Thus, in the third stage, it would be better to only deploy FCUAV to carry out the sensing mission than to deploy any other SUs. In Fig. \ref{fig:SCDAvsbeta}, when the JSC UAV network contains 100 SUs, the maximum UB-ACSA is 8.32 $\times$ $\rm{10^{8}}$ m$\rm{^2}$, which is achieved when there is $\beta_R$ = 0.5430. With the above procedure, we can determine the optimal $\beta_R$ for each UAV given $M$.

\begin{figure}[t]
	\centering
	\includegraphics[width=0.38\textheight]{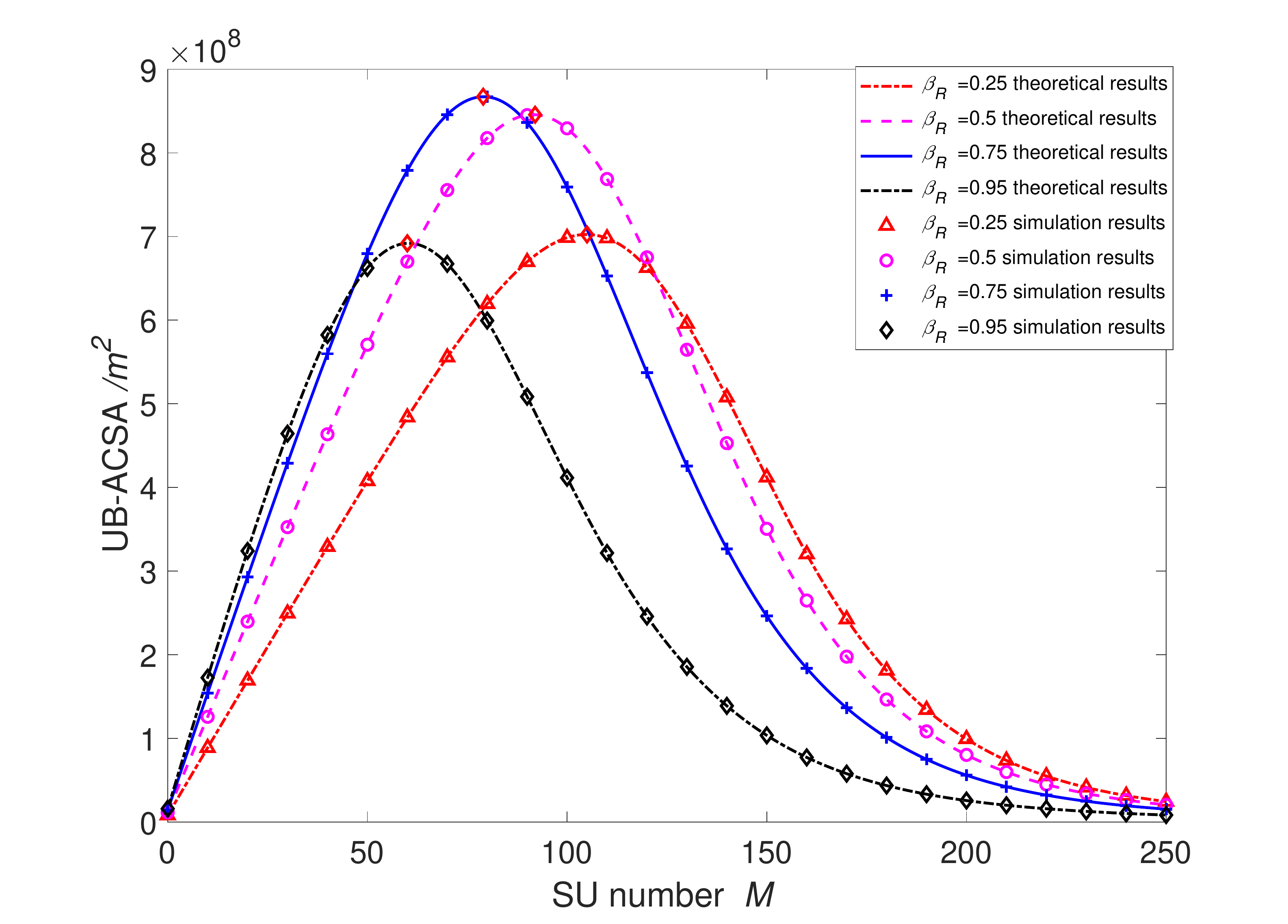}
	\DeclareGraphicsExtensions.
	\caption{UB-ACSA changes against SU number $M$ under different SPR $\beta_R$ when ${K_{h}} =$ 10}
	\label{fig:SCDAvsM}
\end{figure}

\begin{figure}[t]
	\centering
	\includegraphics[width=0.37\textheight]{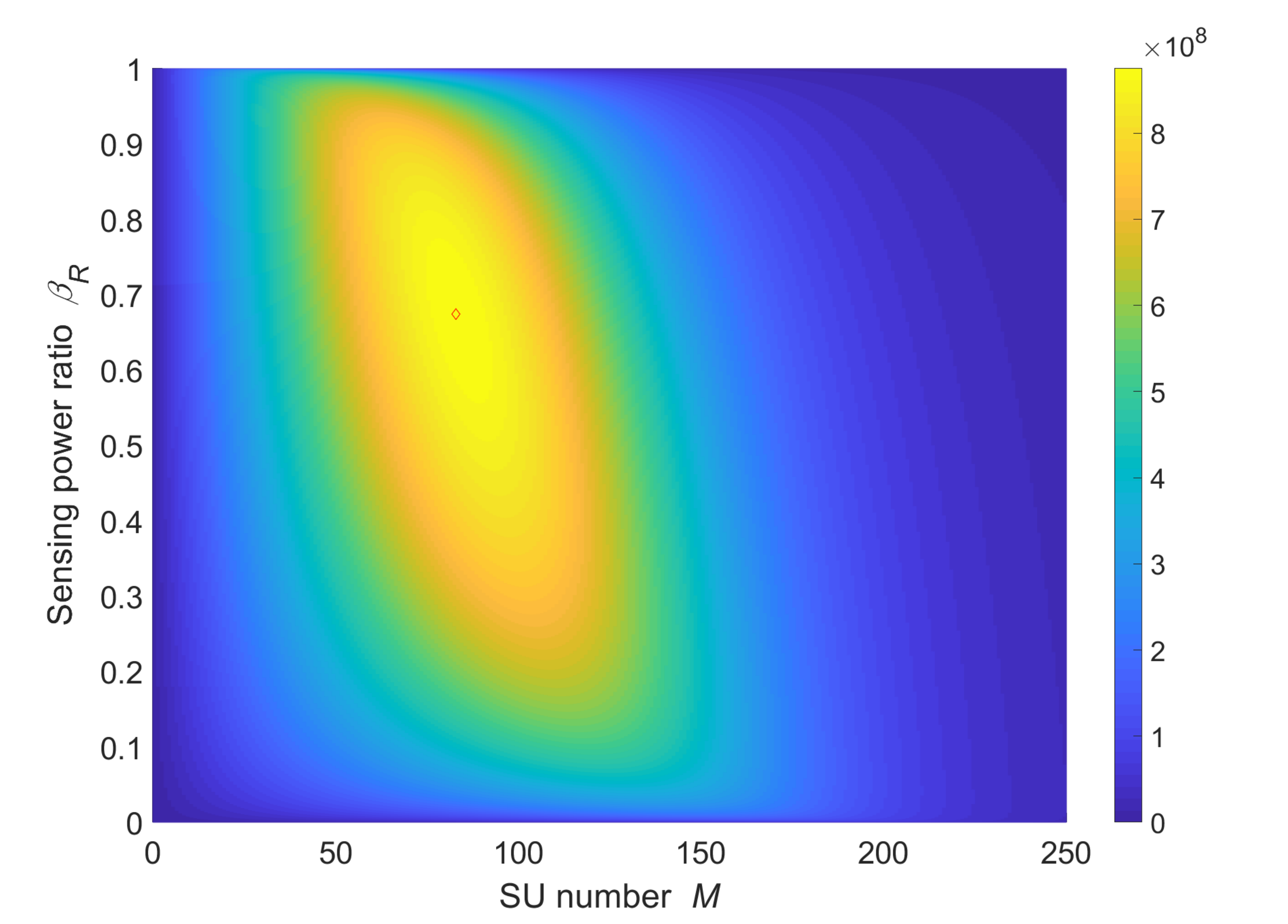}
	\DeclareGraphicsExtensions.
	\caption{ UB-ACSA changes against SPR $\beta_R$ and SU Number $M$ when ${K_{h}} =$ 10}
	\label{fig:SCDA_MBETA}
\end{figure}

Fig. \ref{fig:SCDAvsM} demonstrates the analytical and simulation results of UB-ACSA changing versus $M$ under different $\beta_R$. Each curve of UB-ACSA has three obvious stages. In the first stage where $M$ is relatively small, UB-ACSA increases with the increase of $M$ fast and linearly. This results from that $x_Q$ is considerably large, because the communication capacity requirement for integrating the SUs' sensing data is low. When $M$ gets too large, the second stage comes, where $x_Q$ is much smaller. After reaching the maximum point of UB-ACSA, UB-ACSA decreases fast with the increase of $M$, because $x_Q$ keeps dropping down, $I_{sen}$ gets larger fast, and the overlapped SZs get increasingly larger. If $M$ is so large that $x_Q$ gets extremely small and $I_{sen}$ becomes too large, then $R_d$ converges to 0 m. Therefore, UB-ACSA will become 0 m$^2$ hereafter, which is the third stage. Moreover, fewer SUs are needed to reach the maximum UB-ACSA as $\beta_{R}$ increases, because if the given ComB power decreases, then fewer SUs can be supported to transmit the entire sensing data to FCUAV. Besides, fewer SUs can make $R_d$ be 0 m as $\beta_R$ increases, because the increase of the SenB power and decrease of $x_Q$ lead to the increase of $I_{sen}$. In Fig. \ref{fig:SCDAvsM}, when $\beta_R$ is 0.75, the optimal value of $M$ is 79, which makes UB-ACSA reach the maximum value, i.e., 8.672 $\rm {\times 10^{8}}$ m$\rm ^2$. Using the above procedure, we can easily find the optimal $M$ to accomplish the maximum UB-ACSA, given the value of $\beta_R$.

Fig. \ref{fig:SCDA_MBETA} presents the results of UB-ACSA changing versus $\beta_R$ and $M$. On the one hand, when $\beta_R$ and $M$ are relatively small, the communication capacity is much larger than the requirement for sensing data integration, because the ComB power is considerably large, while the number of UAVs to integrate sensing data is too small. Therefore, the communication capacity for data integration is not completely exploited, i.e., UB-ACSA cannot reach the maximum value when $\beta_R$ and $M$ is small. On the other hand, too large $\beta_R$ will result in the decrease of UB-ACSA because of the decrease of $x_Q$. When $M$ is too large, $x_Q$ also becomes quite small, and $I_{sen}$ will become large enough to make $R_{max}$ decrease to a small value. As a result, when $M$ and $\beta_R$ are too large, UB-ACSA will decrease with the increase of $M$ and $\beta_R$ until UB-ACSA reduces to 0 m$^2$. In Fig. \ref{fig:SCDA_MBETA}, when $\beta_R =$ 0.6741 and $M =$ 83, UB-ACSA reaches the maximum value of 8.748 $\rm{\times 10^8}$ m$\rm{^2}$. In the above procedure, we can easily determine the optimal values of $\beta_R$ and $M$ to achieve the optimal UB-ACSA.



\section{Conclusion}\label{sec:conclusion}
This work has put forward the average cooperative sensing area as performance metric of cooperative sensing in JSC UAV network. To achieve cooperative sensing of UAV network, we propose a novel design of JSC antenna array that consists of the communication subarray and sensing subarray and put forward relative beamforming algorithm. Further, the upper bound of mutual interference of sensing, the minimum required sensing SINR, the maximum sensing range and the maximum cooperative range that can allow SUs to transmit the sensing data to FCUAV successfully are derived. Further, the upper bound of average cooperative sensing area is derived as the tractable performance metric of cooperative sensing, which is related to SPR and the number of UAVs. SPR determines the maximum sensing range, the upper bound of mutual interference of sensing and the maximum cooperative range. The number of SUs determines the maximum cooperative range and the mutual interference of the sensing of UAVs. Finally, the optimal value of SPR and the number of UAVs to achieve the maximum UB-ACSA have been numerically determined. The results can guide the configuration of the JSC UAV cooperative sensing network.

\begin{appendix}
	\begin{theorem} \label{Theo:cumulative density}
		{\rm A 2D uniform point process is generated within a concentric circle area whose inner radius and outer radius are $R_g$ and $x_Q$, respectively. The distance between a generated point and the center is denoted by $R_i$. Let $h$ denote a constant parameter. Then the expectation of $ {{{\left( {{R_i}^2 + 4{h^2}} \right)}^{ - 2}}} $ of ${R_i}$ is}
		\begin{align*}
		E\left\{ {{{\left( {{R_i}^2 + 4{h^2}} \right)}^{ - 2}}} \right\} = \frac{\left[ {{{\left( {{R_g}^2 + 4{h^2}} \right)}^{ - 1}} - {{\left( {{x_Q}^2 + 4{h^2}} \right)}^{ - 1}}} \right]}{{{x_Q}^2 - {R_g}^2}}.
		\end{align*} 
		\begin{proof}	
			{\rm The cumulative density function of $R_i$ should be ${F_{R_i}}\left( r \right) = \frac{{{r^2} - {R_g}^2}}{{{x_Q}^2 - {R_g}^2}} (R_g < r < x_Q )$. Thus, the cumulative distribution function of $Y = {\left( {{{R_i}^2} + 4{h^2}} \right)^{^{ - 2}}}$ ( $h > 0$  and  ${R_g} < {R_i} < {x_Q}$ ) is 
				
				\begin{equation}
				\begin{split}
				{F_Y}\left( y \right) &= \Pr \left( {Y \le y} \right)\\
				&= \Pr \left( {{{R_i}^2} \ge {y^{ - \frac{1}{2}}} - 4{h^2}} \right)\\
				&= \Pr \left( {{R_i} \ge {{\left( {{y^{ - \frac{1}{2}}} - 4{h^2}} \right)}^{\frac{1}{2}}}} \right)\\
				&= 1 - \frac{{\left( {{y^{ - \frac{1}{2}}} - 4{h^2}} \right) - {R_g}^2}}{{{x_Q}^2 - {R_g}^2}},
				\end{split}
				\end{equation}
				where $\left(x_Q^2+4h^2\right)^{-2} \le y\le\left(R_g^2+4h^2\right)^{-2}$.
				Then, the probability density function of $Y$ is 	
				\begin{equation}
				\begin{split}
				{f_{_Y}}\left( y \right) = \frac{{d\left[ {{F_Y}\left( y \right)} \right]}}{{dy}}
				= \frac{1}{{2\left( {{x_Q}^2 - {R_g}^2} \right)}}{y^{ - \frac{3}{2}}},
				\end{split}
				\end{equation}
				where $\left(x_Q^2+4h^2\right)^{-2} \le y\le\left(R_g^2+4h^2\right)^{-2}$.
				Thus, the expectation of $Y$ is	
				\begin{equation}
				\begin{split}
				E\left( Y \right) &= \int_{\left(x_Q^2+4h^2\right)^{-2}}^{{\left(R_g^2+4h^2\right)^{-2}}} {y{f_Y}\left( y \right)dy} \\
				&= \frac{1}{{{x_Q}^2 - {R_g}^2}}\left[ {{{\left( {{R_g}^2 + 4{h^2}} \right)}^{ - 1}} - {{\left( {{x_Q}^2 + 4{h^2}} \right)}^{ - 1}}} \right].
				\end{split}
				\end{equation}}
		\end{proof}
	\end{theorem}
\end{appendix}



%

{\small
	\bibliographystyle{IEEEtran}
	\bibliography{reference}

\begin{thebibliography}{10}
\providecommand{\url}[1]{#1}
\csname url@samestyle\endcsname
\providecommand{\newblock}{\relax}
\providecommand{\bibinfo}[2]{#2}
\providecommand{\BIBentrySTDinterwordspacing}{\spaceskip=0pt\relax}
\providecommand{\BIBentryALTinterwordstretchfactor}{4}
\providecommand{\BIBentryALTinterwordspacing}{\spaceskip=\fontdimen2\font plus
\BIBentryALTinterwordstretchfactor\fontdimen3\font minus
  \fontdimen4\font\relax}
\providecommand{\BIBforeignlanguage}[2]{{%
\expandafter\ifx\csname l@#1\endcsname\relax
\typeout{** WARNING: IEEEtran.bst: No hyphenation pattern has been}%
\typeout{** loaded for the language `#1'. Using the pattern for}%
\typeout{** the default language instead.}%
\else
\language=\csname l@#1\endcsname
\fi
#2}}
\providecommand{\BIBdecl}{\relax}
\BIBdecl

\bibitem{Chen2019Per}
X.~Chen, Z.~Wei, Z.~Fang, H.~Ma, Z.~Feng, and H.~Wu, ``Performance of joint
  radar-communication enabled cooperative {UAV} network,'' \emph{IEEE
  International Conference on Signal, Information and Data Processing 2019},
  pp. 1--4, Dec. 2019.

\bibitem{teacy2010maintaining}
W.~L. Teacy, J.~Nie, S.~McClean, and G.~Parr, ``Maintaining connectivity in
  {UAV} swarm sensing,'' \emph{2010 IEEE Globecom Workshops}, pp. 1771--1776,
  Jan. 2011.

\bibitem{han2013joint}
L.~Han and K.~Wu, ``Joint wireless communication and radar sensing systems-
  state-of-the-art and future prospects,'' \emph{IET Microwaves, Antennas \&
  Propagation}, vol.~7, no.~11, pp. 876--885, Aug. 2013.

\bibitem{Sturm2011Waveform}
C.~Sturm and W.~Wiesbeck, ``Waveform design and signal processing aspects for
  fusion of wireless communications and radar sensing,'' \emph{Proceedings of
  the IEEE}, vol.~99, no.~7, pp. 1236--1259, May. 2011.

\bibitem{gupta2015survey}
A.~Gupta and R.~K. Jha, ``A survey of {5G} network: Architecture and emerging
  technologies,'' \emph{IEEE Access}, vol.~3, pp. 1206--1232, May. 2015.

\bibitem{moghaddasi2016multifunctional}
J.~Moghaddasi and K.~Wu, ``Multifunctional transceiver for future radar sensing
  and radio communicating data-fusion platform,'' \emph{IEEE Access}, vol.~4,
  pp. 818--838, Feb. 2016.

\bibitem{Wu8168372}
J.~{Ellinger}, Z.~{Zhang}, Z.~{Wu}, and M.~C. {Wicks}, ``Dual-use multicarrier
  waveform for radar detection and communication,'' \emph{IEEE Transactions on
  Aerospace and Electronic Systems}, vol.~54, no.~3, pp. 1265--1278, June 2018.

\bibitem{shi2018power}
C.~Shi, F.~Wang, M.~Sellathurai, J.~Zhou, and S.~Salous, ``Power
  minimization-based robust {OFDM} radar waveform design for radar and
  communication systems in coexistence,'' \emph{IEEE Transactions on Signal
  Processing}, vol.~66, no.~5, pp. 1316--1330, Jan. 2018.

\bibitem{liu2018mu}
F.~Liu, C.~Masouros, A.~Li, H.~Sun, and L.~Hanzo, ``{MU-MIMO} communications
  with {MIMO} radar: From co-existence to joint transmission,'' \emph{IEEE
  Transactions on Wireless Communications}, vol.~17, no.~4, pp. 2755--2770,
  Feb. 2018.

\bibitem{mccormick2017simultaneous}
P.~M. McCormick, S.~D. Blunt, and J.~G. Metcalf, ``Simultaneous radar and
  communications emissions from a common aperture, {Part I}: Theory,''
  \emph{2017 IEEE Radar Conference}, pp. 1685--1690, Jun. 2017.

\bibitem{richards2005fundamentals}
M.~A. Richards, ``Fundamentals of radar signal processing,'' \emph{Tata
  McGraw-Hill Education}, 2005.

\bibitem{monzingo2004introduction}
R.~A. Monzingo and T.~W. Miller, ``Introduction to adaptive arrays,''
  \emph{Scitech publishing}, 2004.

\bibitem{frost1972algorithm}
O.~L. Frost, ``An algorithm for linearly constrained adaptive array
  processing,'' \emph{Proceedings of the IEEE}, vol.~60, no.~8, pp. 926--935,
  1972.

\bibitem{shi2005new}
Z.~Shi and Z.~Feng, ``A new array pattern synthesis algorithm using the
  two-step least-squares method,'' \emph{IEEE signal processing letters},
  vol.~12, no.~3, pp. 250--253, Mar. 2005.

\bibitem{skolnik1970radar}
M.~I. Skolnik, ``Radar handbook,'' \emph{McGraw--Hill}, 1970.

\bibitem{zhang2016bio}
W.~M. Z.~W. Zhiping~Zhang, N. Michael~J, ``Bio-inspired {RF} steganography via
  linear chirp radar signals,'' \emph{IEEE Communications Magazine}, vol.~54,
  no.~6, pp. 82--86, June 2016.

\bibitem{stove1992linear}
A.~G. Stove, ``Linear {FMCW} radar techniques,'' \emph{Radar and Signal
  Processing, IEE Proceedings F}, vol. 139, no.~5, pp. 343--350, Oct. 1992.

\bibitem{nathanson1991radar}
F.~E. Nathanson, J.~P. Reilly, and M.~N. Cohen, ``Radar design
  principles-signal processing and the environment,'' \emph{NASA STI/Recon
  Technical Report A}, vol.~91, 1991.

\bibitem{sturm2010performance}
C.~Sturm, T.~Zwick, W.~Wiesbeck, and M.~Braun, ``Performance verification of
  symbol-based {OFDM} radar processing,'' \emph{2010 IEEE Radar Conference},
  pp. 60--63, May 2010.

\bibitem{levy2008principles}
B.~C. Levy, ``Principles of signal detection and parameter estimation,''
  \emph{Springer Science \& Business Media}, 2008.

\bibitem{richards2010principles}
M.~A. Richards, J.~Scheer, W.~A. Holm, and W.~L. Melvin, ``Principles of modern
  radar,'' \emph{Citeseer}, 2010.

\bibitem{mailloux2017phased}
R.~J. Mailloux, \emph{Phased array antenna handbook}.\hskip 1em plus 0.5em
  minus 0.4em\relax Artech house, 2017.

\bibitem{7105651}
Z.~{Zhang}, X.~{Chai}, K.~{Long}, A.~V. {Vasilakos}, and L.~{Hanzo}, ``Full
  duplex techniques for {5G} networks: self-interference cancellation, protocol
  design, and relay selection,'' \emph{IEEE Communications Magazine}, vol.~53,
  no.~5, pp. 128--137, May 2015.

\bibitem{proakis1994communication}
J.~G. Proakis, M.~Salehi, N.~Zhou, and X.~Li, ``Communication systems
  engineering,'' \emph{Prentice Hall New Jersey}, vol.~2, 1994.

\bibitem{molisch2012wireless}
A.~F. Molisch, ``Wireless communications,'' \emph{John Wiley \& Sons}, vol.~34,
  2012.

\bibitem{Yuan8345703}
X.~{Yuan}, Z.~{Feng}, W.~{Xu}, W.~{Ni}, J.~A. {Zhang}, Z.~{Wei}, and R.~P.
  {Liu}, ``Capacity analysis of {UAV} communications: Cases of random
  trajectories,'' \emph{IEEE Transactions on Vehicular Technology}, vol.~67,
  no.~8, pp. 7564--7576, Aug. 2018.

\bibitem{azari2018ultra}
M.~M. Azari, F.~Rosas, K.-C. Chen, and S.~Pollin, ``Ultra reliable {UAV}
  communication using altitude and cooperation diversity,'' \emph{IEEE
  Transactions on Communications}, vol.~66, no.~1, pp. 330--344, Jan. 2018.

\bibitem{goddemeier2015investigation}
N.~Goddemeier and C.~Wietfeld, ``Investigation of air-to-air channel
  characteristics and a {UAV} specific extension to the {Rice} model,''
  \emph{2015 IEEE Globecom Workshops}, pp. 1--5, Dec. 2015.

\bibitem{khuwaja2018survey}
A.~A. Khuwaja, Y.~Chen, N.~Zhao, M.-S. Alouini, and P.~Dobbins, ``A survey of
  channel modeling for {UAV} communications,'' \emph{IEEE Communications
  Surveys $\&$ Tutorials}, Jul. 2018.

\bibitem{Zhalehpour2015Outage}
S.~Zhalehpour, M.~Uysal, O.~A. Dobre, and T.~Ngatched, ``Outage capacity and
  throughput analysis of multiuser {FSO} systems,'' \emph{2015 IEEE 14th
  Canadian Workshop on Information Theory}, Jul. 2015.

\bibitem{Shnidman32133}
D.~A. {Shnidman}, ``The calculation of the probability of detection and the
  generalized marcum {Q}-function,'' \emph{IEEE Transactions on Information
  Theory}, vol.~35, no.~2, pp. 389--400, Mar. 1989.

\bibitem{lakshmanan2007overview}
V.~Lakshmanan, ``Overview of radar data compression,'' \emph{Satellite Data
  Compression, Communications, and Archiving III}, vol. 6683, Sep. 2007.

\end{thebibliography}
}

%
%
%




%

%
%

\ifCLASSOPTIONcaptionsoff
  \newpage
\fi

\end{document}